\documentclass[times, tighten, twocolumn, twocolappendix]{aastex631}

\usepackage{mathrsfs}
\usepackage{amsmath}
\usepackage{rotating}
\usepackage{booktabs}
\usepackage{CJKutf8}
\usepackage{microtype}

\newcommand{\pgpg}{PG~2130$+$099}
\newcommand{\fcs}{$F_{\rm 5100}$}
\newcommand{\ha}{H$\alpha$}
\newcommand{\hb}{H$\beta$}
\newcommand{\hei}{He {\sc i}}
\newcommand{\heii}{He {\sc ii}}
\newcommand{\feii}{Fe {\sc ii}}
\newcommand{\oiii}{[O~{\sc iii}]}
\newcommand{\nimy}{[N~{\sc i}]}
\newcommand{\fevi}{[Fe~{\sc vi}]}
\newcommand{\fevii}{[Fe~{\sc vii}]}

\newcommand{\fhb}{$F_{\rm H\beta}$}

\newcommand{\fhei}{$F_{\rm He~\textsc{i}}$}

\newcommand{\fheii}{$F_{\rm He~\textsc{ii}}$}
\newcommand{\ffeii}{$F_{\rm Fe~\textsc{ii}}$}
\newcommand{\fline}{$F_{\rm line}$}

\newcommand{\fvar}{$F_{\rm var}$}
\newcommand{\rhb}{$R_{\rm H\beta}$}
\newcommand{\rfeii}{$R_{\rm Fe~\textsc{ii}}$}
\newcommand{\rhei}{$R_{\rm He~\textsc{i}}$}

\newcommand{\sigfvar}{$\sigma_{F_{\rm var}}$}

\newcommand{\mbh}{$M_\bullet$}
\newcommand{\mathmdot}{$\dot{\mathscr{M}}$}
\newcommand{\rmax}{$r_{\rm max}$}

\newcommand{\lcs}{$L_{\rm 5100}$}
\newcommand{\rl}{\rhb~$-$~\lcs}
\newcommand*{\difd}{\mathop{}\!\mathrm{d}}

\defcitealias{Hu2020}{I}
\def\kms{km~${\rm s^{-1}}$}

\def\ergs{\rm erg~s$^{-1}$}
\def\ergscm{\rm erg~s$^{-1}$~cm$^{-2}$}
\def\ergscma{\rm erg~s$^{-1}$~cm$^{-2}$~{\AA}$^{-1}$}

\shorttitle{pg2130 draft}
\shortauthors{Zhu-heng Yao et al.}

\def\authormail{
\href{mailto:huc@ihep.ac.cn}{huc@ihep.ac.cn}, \href{mailto:wangjm@ihep.ac.cn}{wangjm@ihep.ac.cn}
}

\begin{document}

\title{Broad-line Region of the Quasar {\pgpg}. II.
Doubling the Size Over Four Years?}

\author[0009-0000-1228-2373]{Zhu-Heng Yao}
\author{Sen Yang}
\affiliation{Key Laboratory for Particle Astrophysics, Institute of High Energy Physics, Chinese Academy of Sciences, 19B Yuquan Road, Beijing 100049, China; \authormail}
\affiliation{School of Physical Science, University of Chinese Academy of Sciences, 19A Yuquan Road, Beijing 100049, People’s Republic of China}

\author[0000-0001-9457-0589]{Wei-Jian Guo}
\affiliation{National Astronomical Observatories, Chinese Academy of Sciences, 20A Datun Road, Chaoyang District, Beijing 100101, China}
\affiliation{Key Laboratory of Optical Astronomy, National Astronomical Observatories, Chinese Academy of Sciences, Beijing 100012, China}

\author[0000-0003-4280-7673]{Yong-Jie Chen}
\affiliation{Key Laboratory for Particle Astrophysics, Institute of High Energy Physics, Chinese Academy of Sciences, 19B Yuquan Road, Beijing 100049, China; \authormail}
\affiliation{Dongguan Neutron Science Center, 1 Zhongziyuan Road, Dongguan 523808, China}

\author[0000-0003-4042-7191]{Yu-Yang Songsheng}
\affiliation{Key Laboratory for Particle Astrophysics, Institute of High Energy Physics, Chinese Academy of Sciences, 19B Yuquan Road, Beijing 100049, China; \authormail}

\author[0000-0003-2024-1648]{Dong-Wei Bao}
\affiliation{National Astronomical Observatories, Chinese Academy of Sciences, 20A Datun Road, Chaoyang District, Beijing 100101, China}

\author[0000-0003-3825-0710]{Bo-Wei Jiang}
\author{Yi-Lin Wang}
\author[0000-0002-5595-0447]{Hao Zhang}
\affiliation{Key Laboratory for Particle Astrophysics, Institute of High Energy Physics, Chinese Academy of Sciences, 19B Yuquan Road, Beijing 100049, China; \authormail}
\affiliation{School of Physical Science, University of Chinese Academy of Sciences, 19A Yuquan Road, Beijing 100049, People’s Republic of China}

\author{Chen Hu}
\author[0000-0001-5841-9179]{Yan-Rong Li}
\author[0000-0002-5830-3544]{Pu Du}
\author[0000-0001-5981-6440]{Ming Xiao}
\affiliation{Key Laboratory for Particle Astrophysics, Institute of High Energy Physics, Chinese Academy of Sciences, 19B Yuquan Road, Beijing 100049, China; \authormail}

\author{Jin-Ming Bai}
\affiliation{Yunnan Observatories, The Chinese Academy of Sciences, Kunming 650011, People’s Republic of China}

\author[0000-0001-6947-5846]{Luis C. Ho}
\affiliation{Department of Astronomy, School of Physics, Peking University, Beijing 100871, China}
\affiliation{Physics Department, Nanjing Normal University, Nanjing 210097, China}

\author[0000-0002-1207-0909]{Michael S. Brotherton}
\affiliation{Department of Physics and Astronomy, University of Wyoming, Laramie, WY 82071, USA}

\author[0000-0003-0487-1105]{Jes\'us Aceituno}
\affiliation{Centro Astronomico Hispano Alem\'an, Sierra de los filabres sn, 04550 gergal. Almer\'ia, Spain}
\affiliation{Instituto de Astrof\'isica de Andaluc\'ia, Glorieta de la astronom\'ia sn, 18008 Granada, Spain}

\author[0000-0003-2662-0526]{Hartmut Winkler}
\affiliation{Department of Physics, University of Johannesburg, PO Box 524, 2006 Auckland Park, South Africa}

\author[0000-0001-9449-9268]{Jian-Min Wang}
\affiliation{Key Laboratory for Particle Astrophysics, Institute of High Energy Physics, Chinese Academy of Sciences, 19B Yuquan Road, Beijing 100049, China; \authormail}
\affil{University of Chinese Academy of Sciences, 19A Yuquan Road, Beijing 100049,  China}
\affil{National Astronomical Observatories of China, Chinese Academy of Sciences, 20A Datun Road, Beijing 100020, China}

\collaboration{99}{(Seambh Collaboration)}

\begin{abstract}
Over the past three decades, multiple reverberation mapping (RM) campaigns conducted for the quasar {\pgpg} have exhibited inconsistent findings with time delays ranging from $\sim$10 to $\sim$200 days.
To achieve a comprehensive understanding of the geometry and dynamics of the broad-line region (BLR) in {\pgpg}, we continued an ongoing high-cadence RM monitoring campaign using the Calar Alto Observatory 2.2m optical telescope for an extra four years from 2019 to 2022.
We measured the time lags of several broad emission lines (including {\heii}, {\hei}, {\hb}, and {\feii}) with respect to the 5100~{\AA} continuum, and their time lags continuously vary through the years.
Especially, the {\hb} time lags exhibited approximately a factor of two increase in the last two years.
Additionally, the velocity-resolved time delays of the broad {\hb} emission line reveal a back-and-forth change between signs of virial motion and inflow in the BLR.
The combination of negligible ($\sim$10\%) continuum change and substantial time-lag variation (over two times) results in significant scatter in the intrinsic {\rl} relationship for {\pgpg}.
Taking into account the consistent changes in the continuum variability time scale and the size of the BLR, we tentatively propose that the changes in the measurement of the BLR size may be affected by ``geometric dilution''.
\end{abstract}


\section{Introduction} \label{sec:intro}
In recent years, the existence of supermassive black holes (SMBHs) has received significant observational and theoretical support~\citep{Kormendy1995}, highlighting their indispensable contribution to our understanding of cosmic evolution.
The masses of SMBHs are often correlated with the properties of their host galaxies, implying their crucial contribution to the evolution of galactic systems~\citep{Kormendy2013, Ho2008}.
Reverberation mapping (RM) is a well-established methodology to estimate the masses of SMBHs, which also allows the simultaneous investigation of the geometry and kinematics of the surrounding SMBH environment, particularly the broad-line region (BLR) of active galactic nuclei (AGNs), through the measurement of the broad emission line's response relative to the continuum (e.g.,~\citealt{Blandford1982, Peterson1993, Peterson1998a, Peterson2004, Shen2015, Du2018, Czerny2019, Zajacek2021, Bao2022, Woo2024, Zastrocky2024, Li2024a}).

As a potential candidate super-Eddington accretor, the bright local quasar, {\pgpg} ($V\sim$ 14.3 mag, $z=$ 0.063)\footnote{These values are given by the NASA/IPAC Extragalactic Database\\ (NED; \url{http://ned.ipac.caltech.edu/}).}, was observed several times in RM studies.
The initial RM observation of this specific quasar was conducted by \citet{Kaspi2000}, commencing in March 1991 and extending over a span of 7.5 years, and employed the Wise Observatory 1m telescope and the Steward Observatory 2.3m telescope.
In their study, they reported a time delay of 188$_{-27}^{+136}$ days for the broad {\hb} emission line and identified {\pgpg} as an outlier of the {\rl} relation, significantly stimulating subsequent observational investigations into this specific source.
However, their findings would be affected by the sparse sampling, which would have precluded detecting time delays shorter than the cadence.
\citet{Grier2008} reanalyzed these data and obtained shorter time delays of 12.2$_{-2.5}^{+1.5}$, 13.7$_{-0.8}^{+1.8}$, and 46.5$_{-2.9}^{+4.0}$ days for the years 1993-1995 respectively (see Table~5 in \citealt{Grier2008}), which differ significantly from the lag results of the overall three-year (203.5$_{-17.4}^{+5.2}$ days) and the 7.5-year measurements.

\citet{Grier2008} also observed 21 epochs from September to December 2007 using the MDM~2.4m telescope and measured a time delay of 22.9$_{-4.3}^{+4.4}$ days.
Although the results suggested that {\pgpg} may again ``conform to the {\rl} relation'', the short duration and non-uniform sampling of this campaign make this conclusion less convincing.

Therefore, \citet{Grier2012} conducted more observations from August to December 2010 using the MDM Observatory 1.3m McGraw-Hill telescope and the Crimean Astrophysical Observatory 2.6m Shajn telescope, collecting data of 88 epochs from 120 nights.
They measured the {\hb} time lag as 9.6$\pm1.2$ days from the cross-correlation function (CCF), and 12.8$_{-0.9}^{+1.2}$ days from the stochastic process estimation for AGN reverberation (SPEAR; \citealt{Zu2011}).
Both results pinpointed {\pgpg} as a noticeable outlier of the {\rl} relationship.
However, \citet{Grier2013a} used the maximum-entropy method to reconstruct the velocity-delay maps of the data and determined a time delay of $\sim$31 days, which was deemed to be the best estimate yet by \citet{Bentz2013}.
They explained that the CCF/SPEAR measurements in \citet{Grier2012} deviated due to missing key epochs with variability characteristics during the 2010 campaign.

From June 2017 to January 2019, \citet{Hu2020} (hereafter Paper~\citetalias{Hu2020}) measured time delays of the broad {\feii}, {\hb}, {\hei} and {\heii} emission lines of {\pgpg} using the Centro Astron\'omico Hispano-Alem\'an (CAHA) 2.2m telescope in Spain.
The geometry and kinematics of the BLR exhibited perplexing changes over this two-year period.
Notably, the {\hb} time lag remained relatively stable: 22.6$_{-3.6}^{+2.7}$ days in the first year (2017) and 27.8$_{-2.9}^{+2.9}$ days in the second year (2018); but the velocity-resolved kinematic structure transitioned from the Keplerian motion in 2017 to the inflow in 2018.
Another interesting phenomenon noted was that changes of the {\hei} and {\feii} lags in 2018 disrupted the stratified onion structure previously revealed in 2017.
Furthermore, there was no significant difference between the mean spectra in the two years, but the RMS spectra and line responses were very different.
All the changes occurred in less than one year, which is much smaller than the dynamical timescale ($\sim$25 light-days / 2000 {\kms} $\simeq$ 10.3 yrs) of the BLR variations for {\pgpg}.
These unresolved issues underscore the importance of continued long-term and high-cadence RM monitoring of {\pgpg} to elucidate the geometry, kinematics, and potential evolution of the BLR.

Based on the findings of Paper~\citetalias{Hu2020}, we have continued the RM monitoring for {\pgpg} with the CAHA 2.2m telescope from June 2019 to January 2023, and our results are presented here as follows.
The data acquisitions and light curve measurements are described in Sections~\ref{sec:obsdata} and ~\ref{sec:lc}, respectively.
Section~\ref{sec:analysis} analyzes the characteristics of light curves, geometry, and kinematics of BLR.
In Section~\ref{sec:discussion}, we determine the {\rl} relationship based on all RM results, discuss the calculations of black hole mass and accretion rate, and consider several factors that may affect the time lag measurements.
All results are summarized in Section~\ref{sec:summary}.
In this work, we assume a cosmology with $H_0=$ 70~\kms~Mpc$^{-1}$ and $\Omega_{\rm M}=$ 0.3.

\section{Observation and Data Reduction} \label{sec:obsdata}
The SEAMBH collaboration started a long-term RM monitoring project (hereafter CAHA-SEAMBH project) in May 2017 focused on PG targets with high accretion rates (Paper~\citetalias{Hu2020}, \citealt{Hu2020a, Hu2021, Donnan2023}), mainly carried out at the 2.2-meter telescope in the Calar Alto Observatory.
As one of the brightest PG quasars with numerous previous studies, {\pgpg} was monitored with high priority between June 2017 and January 2023 with 330 epochs observed under the CAHA-SEAMBH project (219 epochs between June 2019 and January 2023).
For each epoch, optical photometry was first carried out using a Johnson $V$ filter.
It was followed by spectroscopy of the AGN, where a nearby comparison star with near-constant flux was observed simultaneously to ensure accurate flux measurements \citep{Kaspi2000, Du2014, Hu2020}.
The magnitude of the comparison star given by the Sloan Digital Sky Survey (SDSS; \citealt{York2000}) is $g=$ 15.1.
The photometry and spectroscopy calibration were performed using standard IRAF v2.16~\citep{Tody1986, Tody1993} and Python-based procedures.

\subsection{Photometry} \label{subsec:phot}
\begin{figure*}[!ht]
    \includegraphics[width=2\columnwidth, trim=0 0 60 0]{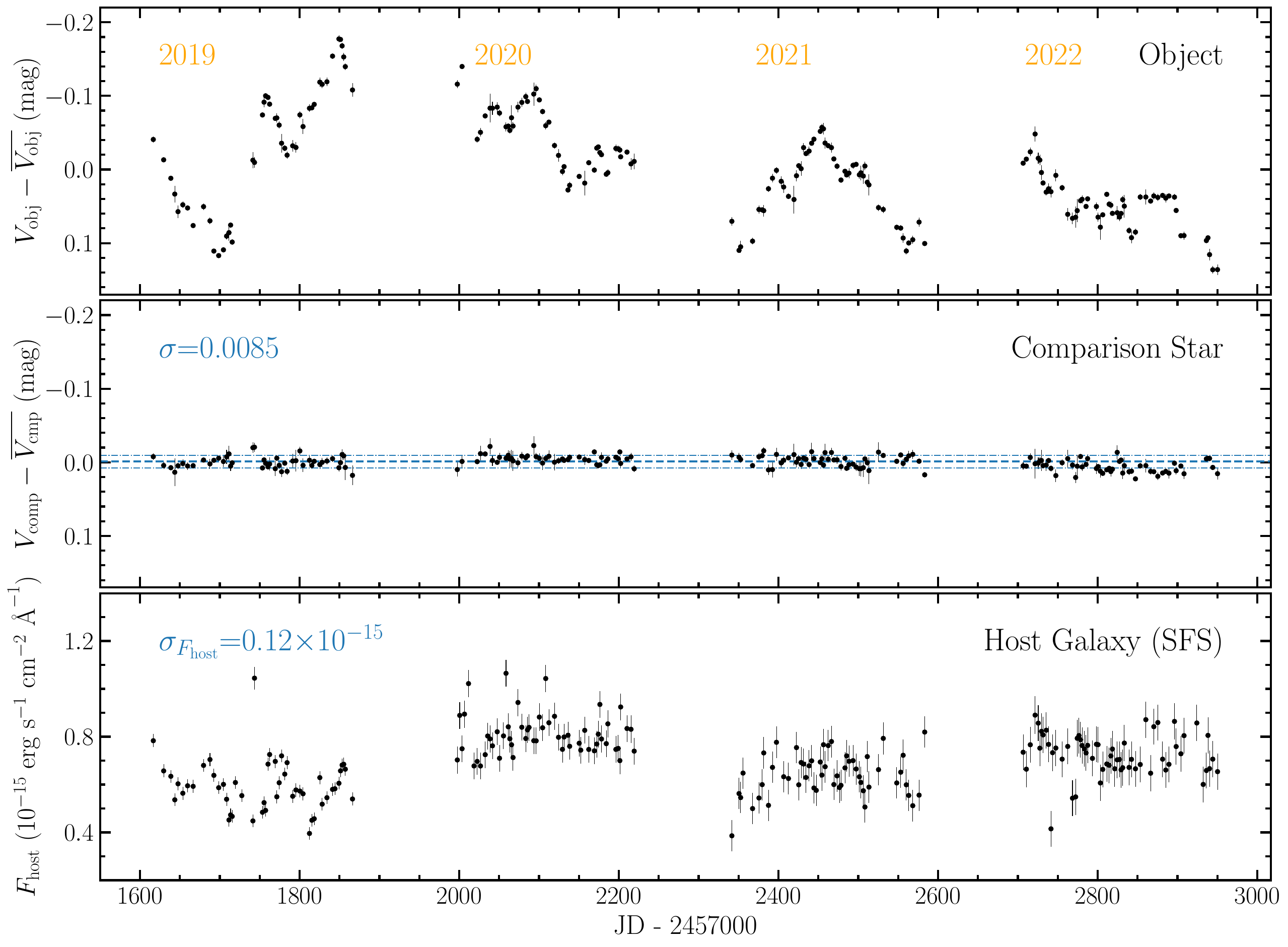}
    \caption{
    Photometric light curves in the $V$-band for {\pgpg} (top) and its comparison star (middle) in units of differential magnitudes relative to reference stars, with mean values subtracted respectively, and light curve of the host galaxy obtained by SFS (bottom).
    Different years are annotated using orange in the top panel.
    The scatter of the comparison star light curve is 0.0085 mag, implying that the accuracy of our flux calibration is better than 1\%.
    The scatter of the host starlight, $\sigma_{F_{\rm host}}\sim~0.12\times10^{-15}$~\ergscma, accounts for only $\sim$2\% of the mean 5100~{\AA} flux, indicating that the host galaxy's impact is minimal.
    }
    \label{fig:photlc}
\end{figure*}

In each individual night, at least three $V$ band images with 20-seconds exposure each were obtained and combined thereafter to ensure photometric accuracy.
IRAF was deployed to subtract the bias and the flat field.
For our aperture photometry we set the radius of the source aperture as 2$\farcs$65 and used the mean flux inside the annulus between radii of 5$\farcs$3 and 7$\farcs$95 as the sky background.
The relative photometric light curves were obtained from differential photometry using 14 nearby reference stars with no obvious flux variation within the field of 8$\farcm$8 $\times$ 8$\farcm$8.
The photometric light curves of {\pgpg} and its comparison star are shown in Figure~\ref{fig:photlc}.
The comparison star exhibits a scatter of $\sim$0.01 mag, which provides convincing evidence that it is suitable for conducting spectroscopic flux calibration.

\subsection{Spectroscopy} \label{subsec:spec}
The RM spectroscopy has been carried out with the Calar Alto Faint Object Spectrograph (CAFOS), which is equipped with a variety of grisms and a width-adjustable slit.
For {\pgpg}, we use a combination of a 3{\arcsec} wide slit (wider than the typical seeing of 1$''$--2$''$) and the G-200 grism, which covers the wavelength range 4000--8500~{\AA} in the observed frame with a dispersion of 4.47~{\AA}~pixel$^{-1}$.
Normally two 600s-exposure spectra were taken each night.
Standard calibration frames (bias, dome flats, wavelength calibration lamps of HgCd/He/Rb and spectrophotometric standard stars) were usually obtained on the same night (if weather and time permits).

For spectral data reduction, we followed the standard IRAF processes, including bias-removal, flat-fielding, and wavelength calibration.
The target spectra were extracted from a uniform wide aperture of 10$\farcs$6 to avoid light losses \citep{Peterson1995}, with a background region of 7$\farcs$95--15$\farcs$9 on either side.
As illustrated in the bottom panel of Figure~\ref{fig:photlc} and calculated in Section~\ref{sec:lc}, the host contribution is just $\sim$14\% to the total flux, and the scatter of the host starlight ($\sigma_{F_{\rm host}}\sim0.12\times10^{-15}$~\ergscma) equals only $\sim$2\% of the mean 5100~{\AA} flux, indicating that the host galaxy's impact is minimal.
The absolute flux calibration was performed as follows:
(1) generating a fiducial spectrum of the comparison star by combining several spectra taken under good weather conditions;
(2) comparing the fiducial spectrum with each individual spectrum of the comparison star to generate a sensitivity function;
(3) calibrating the flux of the object by applying the corresponding sensitivity function to the spectra of the targets. 

\begin{figure*}[!ht]
    \includegraphics[width=2\columnwidth, trim=-10 0 50 0]{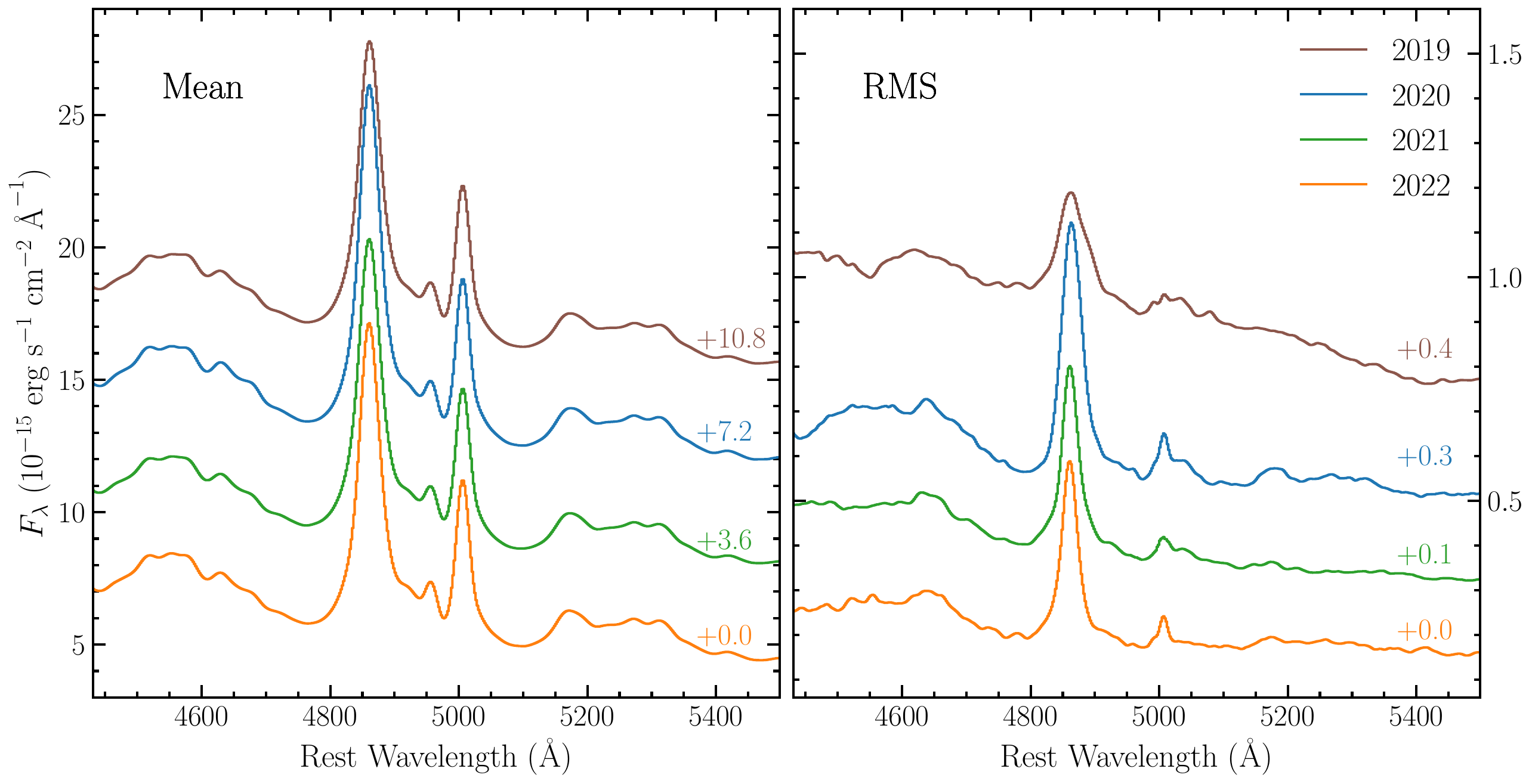}
    \caption{Mean and RMS spectra for each year.
    The spectra are shifted for clarity, and the values of the shifts are indicated on the corresponding spectra with the respective color-coded numbers.
    The legend in the right panel marks the years 2019-2022 in different colors.}
    \label{fig:mr}
\end{figure*}

Spectra with poor signal-to-noise ratio (S/N) were removed and those from the same night were combined for measurements and analysis.
The typical S/N per pixel at 5100~{\AA} of our spectra is $\sim$86.
We then calculate the mean and root mean square (RMS) spectra for each year following the definitions \citep{Peterson2004}:
\begin{equation}
    F_{\lambda}=\frac{1}{N}\sum_{i=1}^{N}F^{i}_{\lambda}, \label{eq:mean}
\end{equation}
\begin{equation}
    S_{\lambda}^2=\frac{1}{N-1}\sum_{i=1}^{N}(F^{i}_{\lambda}-\overline{F_{\lambda}})^2, \label{eq:rms}
\end{equation}
where $N$ is the number of epochs (with an associated spectrum) in one year, and $F^{i}_{\lambda}$ represents the $i$-th spectrum.
Figure~\ref{fig:mr} shows the mean and RMS spectra for 2019 to 2022.
It is evident that the mean spectra exhibit a high degree of similarity, whereas the RMS spectra display striking dissimilarities, specifically for the Balmer lines.
In year 2019, the RMS spectrum is similar to that of 2017 (see Figure~2 in Paper~\citetalias{Hu2020}), but the {\hb} profile is significantly broader than the {\hb} profiles of RMS seen in the last three years.
Notably, the features in the RMS spectra around $\sim$5010~{\AA} exhibit some minor variabilities, which are most likely attributed to varying flux of {\oiii} due to varying part of the extended {\oiii}-emission region in the slit caused by varying seeing or centering.
Since we do not employ {\oiii} for flux calibration, these residuals have not effect on our calibration.

\section{Light Curve Measurements} \label{sec:lc}
For further investigation of the geometry and kinematics with potential evolution in BLR of {\pgpg}, the light curve has been split into segments by the observational gaps.
The seasonal time spans and mean sampling interval can be found in Table~\ref{tab:obs}.

\begin{deluxetable*}{cccccc}
\setlength{\tabcolsep}{26pt} 
\tablecolumns{6}
\tabletypesize{\footnotesize}
\tabcaption{Statistical Analysis of Observations of {\pgpg} from the CAHA 2.2m Telescope \label{tab:obs}}
\colnumbers
\tablehead{
\colhead{Year} &
\colhead{JD - 2457000} &
\colhead{Dates} &
\colhead{Period} &
\colhead{Epochs} &
\colhead{$\Delta{T}$} \\
&& (YY.MM.DD) & (day) &  & (day) 
}
\startdata
2019 & 1616.67 - 1866.27 & 19.05.12 - 20.01.17 & 250 & 50 & 5.0 \\
2020 & 1997.63 - 2219.27 & 20.05.27 - 21.01.04 & 222 & 55 & 4.0 \\
2021 & 2341.65 - 2583.30 & 21.05.06 - 22.01.03 & 242 & 53 & 4.6 \\
2022 & 2706.68 - 2950.27 & 22.05.06 - 23.01.05 & 244 & 61 & 4.0 \\
\hline
All  & 1616.67 - 2950.27 & 19.05.12 - 23.01.05 & 958 & 219 & 4.4 \\
\enddata
\tablecomments{Column (1) denotes the year of spectroscopic monitoring, which refers to the annual season beginning in May of the given year and going up to January of the following year. Columns (2) and (3) list the start and end dates and corresponding Julian Dates (JD), respectively. Columns (4)-(6) quantify the observation period, sampling epochs, and mean sampling interval of each year.}
\end{deluxetable*}

\vspace{-8mm}
Before further analyzing the light curves, we employed an extinction law with $R_V=$ 3.1 \citep{Cardelli1989}, and adopted a $V$-band extinction of 0.122 mag obtained from the NED \citep{Schlafly2011} to correct for Galactic reddening.
We also adjusted the spectra to their rest frame adopting $z=$ 0.063.

We employ two distinct methodologies to measure the light curves.
Typically, most conventional RM studies have utilized a straightforward total line integration approach (referred to here as the integration scheme) to compute the flux for individual, non-blended emission lines, such as {\hb} and {\hei}.
For highly blended emission lines like {\feii} and {\heii} emissions, the above methodology however proves to be less robust.
In such cases, it is better to first decompose the blended components using suitable physical assumptions and to use the resulting component flux densities for the light curves; we refer to this as the spectral fitting scheme (SFS; e.g., Paper~\citetalias{Hu2020}).
Therefore, in this work, we quantify the light curves of {\hb} and {\hei} using the total line strength, while the SFS is employed to extract the {\feii} and {\heii} measurements.

\subsection{Integration Scheme} \label{subsec:intg}
\begin{figure*}[!ht]
    \includegraphics[width=2\columnwidth, trim=5 0 75 0]{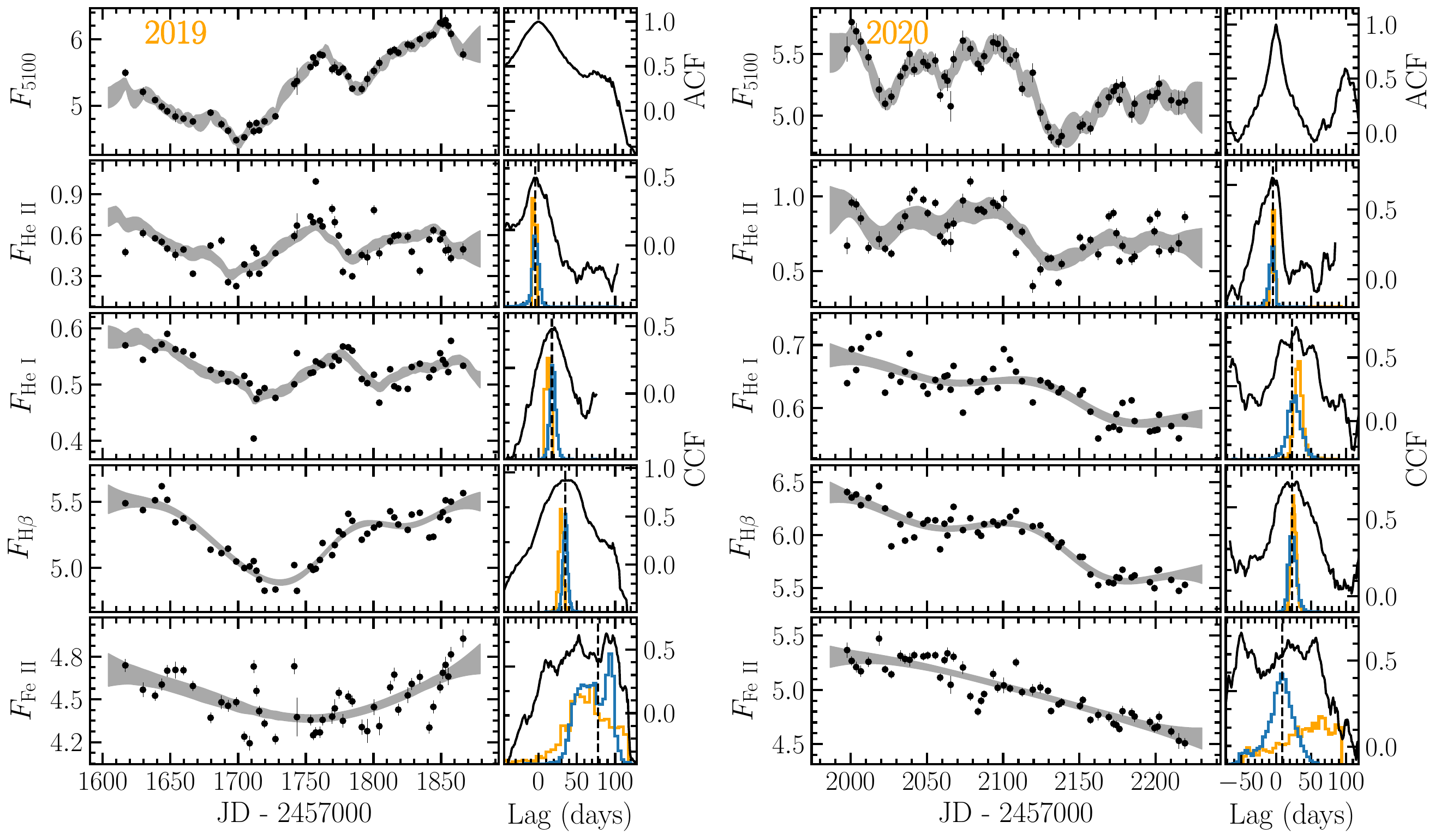}
    \caption{
    Light curves and the results of time series analysis for 2019 (left) and 2020 (right).
    In each case, the left panels from top to bottom show the light curves of the AGN continuum at 5100~{\AA} and the broad {\heii}, {\hei}, {\hb} and {\feii} lines (the fluxes of {\hei} and {\hb} lines are from the integration scheme, while {\heii} and {\feii} from the SFS; see more details in Section~\ref{sec:lc}).
    The light curves filled with grey are the reconstructed light curves by MICA.
    The units of the fluxes of continuum and emission lines are 10$^{-15}$~{\ergscma} and 10$^{-13}$~{\ergscm} respectively.
    The right panels from top to bottom separately correspond to the ACF of the AGN continuum and the CCFs between emission lines and continuum (black solid lines).
    The black dashed lines represent the medians of lags from CCFs, while the time delay distributions measured by different methods are indicated with different colors: blue histograms for CCCDs and orange for MICA results.
    }
    \label{fig:ccfmica34}
\end{figure*}

\begin{figure*}[!ht]
    \includegraphics[width=2\columnwidth, trim=5 0 75 0]{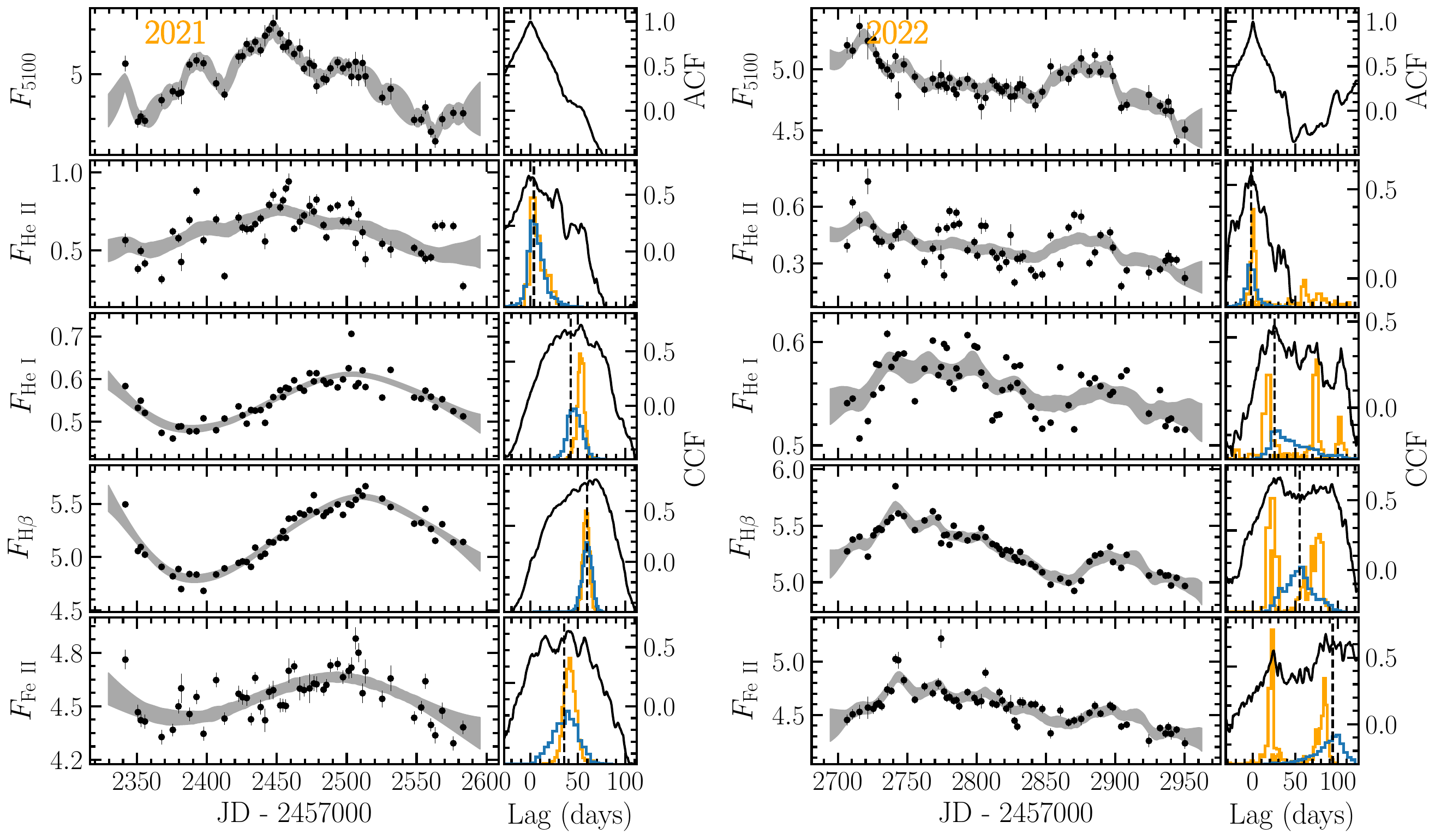}
    \caption{
    Same as Figure~\ref{fig:ccfmica34}, but for 2021 (left) and 2022 (right).
    }
    \label{fig:ccfmica56}
\end{figure*}

The integration window for the {\hb} emission line was defined as 4810--4910~{\AA}, while for the {\hei} emission line, it was set as 5825--5925~{\AA}.
For each line, we fitted a straight-line continuum by using two selected windows on the redward and blueward of the emission.
The average flux densities within the redward continuum window of {\hb} emission, i.e. the range of 5085--5115~{\AA}, were determined to correspond to the 5100~{\AA} continuum.
The resulting flux densities of the emission line were obtained by subtracting the underlying continuum flux assumed to correspond to a straight-line between the blue and red continuum windows.
The windows for the emission lines and corresponding continuum can be seen in Table~\ref{tab:wvwins} and Figure~\ref{fig:mr}.
The light curves of the {\hb}, {\hei} lines and the 5100~{\AA} continuum are displayed in Figure~\ref{fig:ccfmica34} and \ref{fig:ccfmica56} for each year.

\begin{deluxetable}{cccl}
\setlength{\tabcolsep}{15pt} 
\tablecaption{Rest-frame wavelength windows used for measuring the emission-line and continuum fluxes.\label{tab:wvwins}}
\tablecolumns{4}
\tablewidth{\textwidth}
\tablehead{
\colhead{Line} & \colhead{Emission} 
& \multicolumn{2}{c}{Continuum} 
\\\cline{3-4}
&& \colhead{Blueward } & \colhead{Redward} \\
& \colhead{({\AA})} & \colhead{({\AA})} & \colhead{({\AA})} 
}
\startdata
{\hb} & 4810-4910 & 4740-4790 & $\raggedright$ 5085-5115 \tablenotemark{\textsuperscript{a}} \\
{\hei} & 5825-5925 & 5750-5800 & $\raggedright$ 5950-6000 \\
\enddata
\tablenotetext{\textsuperscript{a}}{The integration wavelength window for the 5100~{\AA} continuum.}
\end{deluxetable}

\vspace{-4mm}
\subsection{Spectral Fitting Scheme} \label{subsec:fit}
The SFS employed in this paper employs the treatment used in \citet{Hu2012, Hu2015} and Paper~\citetalias{Hu2020}, involving the simultaneous fittings of multiple spectral components using the Levenberg-Marquardt method to minimize $\chi^{2}$.
The components utilized in our SFS are:
(1) a single power law for the AGN continuum;
(2) a template from \citet{Boroson1992} convolved with a Gaussian function for {\feii} emission lines;
(3) a template with 11 Gyr age and metallicity $Z=$ 0.05 from \citet{Bruzual2003} for host galaxy starlight;
(4) a double Gaussian for the {\hb} emission line;
(5) a single Gaussian for broad {\heii} emission;
(6) a single Gaussian employed for both {\oiii}~$\lambda\lambda$4959,~5007 lines, with identical widths and velocities, and constrained to have a flux ratio of 1/3;
(7) a single Gaussian for narrow {\heii} emission, with the same width and velocity as {\oiii};
(8) a single Gaussian with the same width and velocity shift as {\oiii} for the coronal lines {\fevii}~$\lambda$5158, {\fevi}~$\lambda$5176 and {\nimy}~$\lambda$5199.
Due to its weak intensity, constituting only approximately 2\% of the total {\hb} flux, the narrow component of the {\hb} emission is not included in the SFS analysis (Paper~\citetalias{Hu2020}).
It is noteworthy that our fitting model includes some simplifications:
For instance, we have adopted a single stellar population template for the host galaxy, and some weak emission lines are not considered.
However, they do not affect the measurements of emission lines in this work.

Decomposing the spectra using SFS allows us to directly extract the flux, width, and shift values of various emission lines.
The complex blending of the {\heii} emission line with {\feii} in individual-night and mean spectra poses a particular challenge for accurate fitting.
However, as illustrated in Figure~\ref{fig:mr}, the {\heii} emission is visibly prominent in each RMS spectrum, offering a potential solution to mitigate this issue.
To do so, we employ a two-step approach.
Initially, we perform fitting on the RMS spectra and extract the best-fit parameters (width and shift) of the {\heii} line.
These parameters are then fixed for the purpose of fitting and determining the fluxes of the {\heii} emission in the mean and individual-night spectra.
An additional advantage of first fitting of each RMS spectrum is that we exclude all narrow line components (e.g., {\oiii}, narrow {\heii} and several coronal lines), whose fluxes remain unchanged.
It is worth mentioning that the {\heii} profiles in the single-epoch spectra are not necessarily the same as that in the RMS spectrum, so the {\heii} fluxes we obtained may not be the total flux.
However, by fixing the line widths and shifts of {\heii} in SFS to those in the RMS spectrum, what we measured could be the variable part of the {\heii} fluxes, thus, the measured time lags may not be affected much by this simplification.
Therefore, the components (1)-(5) listed previously are the only ones utilized in the fitting of the RMS spectra.
Subsequently, after fitting the RMS spectrum for each year, we proceed to employ all 8 components listed earlier to fit the mean spectra and all individual-night spectra, with the width and shift of {\heii} fixed to the best fit parameters obtained from the fittings of the RMS spectra.

In addition to the constraints on the width and velocity shift of the {\heii} line, we also apply further constraints to specific components during the fitting process of individual-night spectra: 
(a) The slope of the AGN continuum is fixed to the best-fit value of the mean spectrum \citep{Hu2015};
(b) The velocity width and shift of the host galaxy contribution are fixed to the corresponding values of {\oiii}~$\lambda$5007 from the best-fit result of the mean spectrum for the respective year (Paper~\citetalias{Hu2020});
(c) For the three coronal lines, we constrain their velocity widths and shifts to those of {\oiii}~$\lambda$5007, while maintaining their flux ratios relative to {\oiii} as determined from the best-fit measured for the mean spectrum \citep{Hu2015}.
For completeness and improving the precision of the fitting scheme, we also attempt to restrict the profile of the {\feii} emission by fixing its width and shift to the best-fit values obtained from the mean spectrum, which has negligible impact on the flux measurement of the {\feii} emission line.
In total, the fitting of each individual-night spectrum involves 35 parameters, of which 19 are fixed and 16 are allowed to vary freely.

\begin{figure}[!ht]
    \includegraphics[width=\columnwidth, trim=20 0 0 0]{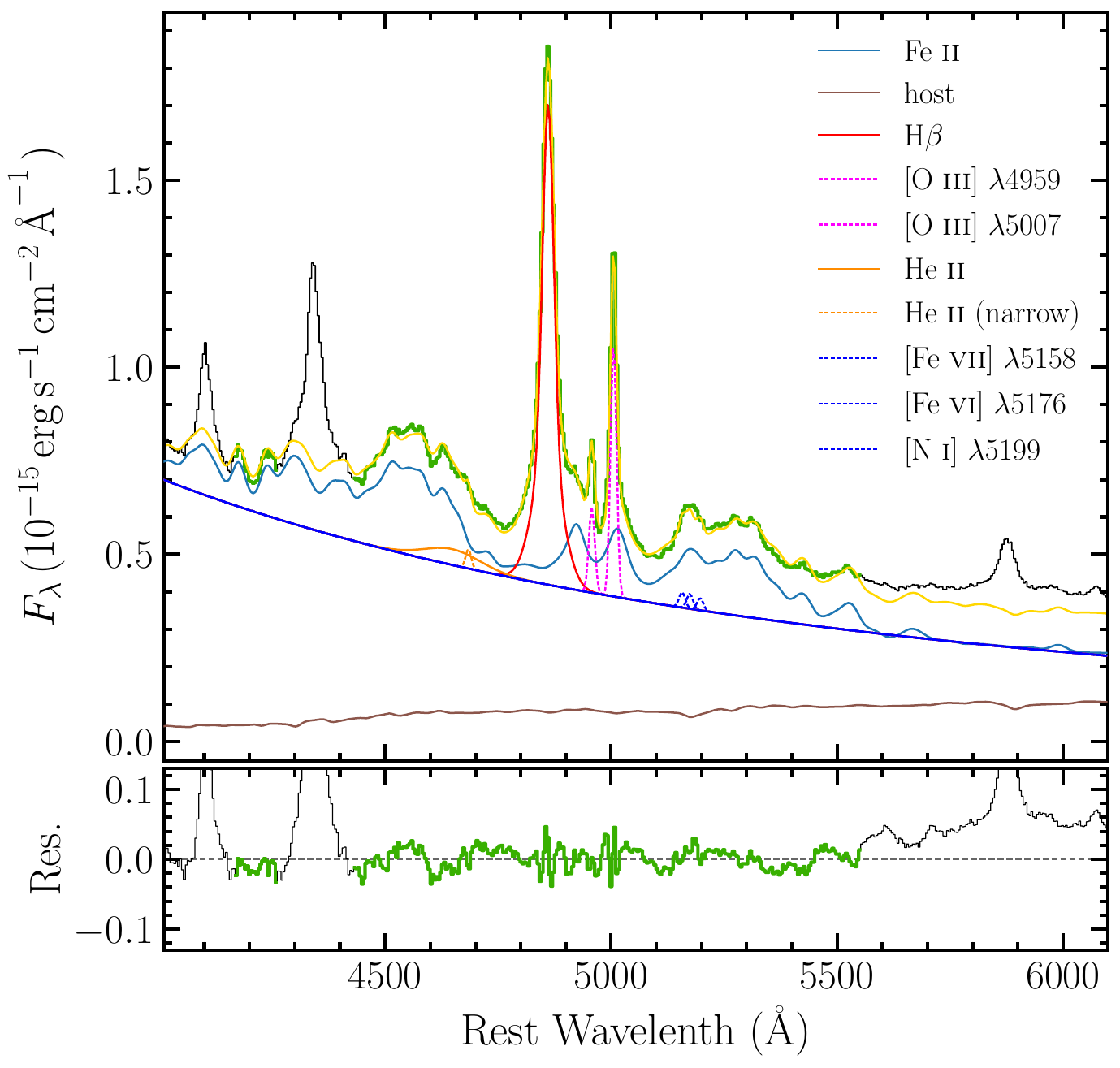}
    \caption{
    An illustrative example of the SFS for an individual-night spectrum of {\pgpg} (JD 2459483.399), spanning the wavelength range of 4150--5550~{\AA}, with the exclusion of the narrow $\rm H\gamma$ window from 4260--4430~{\AA}.
    The top panel showcases the spectrum, with fitted pixels depicted in green and excluded pixels in black, alongside the corresponding best-fit model represented in yellow.
    As lined out in Section~\ref{subsec:fit}, the spectrum consists of several components: an AGN power-law continuum (blue), host galaxy (sky blue), a set of {\feii} emissions (blue), broad {\hb} (red), broad {\heii} (orange), and a set of narrow emission lines (dashed line in various colors).
    The residuals are presented in the bottom panel (green for fitted pixels and black for excluded), providing a visual representation of the deviations between the observed spectrum and the best-fit model.
    }
    \label{fig:fitexample}
\end{figure}

We conduct the SFS by using DASpec\footnote{DASpec is available at: \url{https://github.com/PuDu-Astro/DASpec}} \citep{daspec_dupu}, a specialized software designed for multi-component spectral fitting of AGNs.
All the fittings are performed within the wavelength range of 4150--5550~{\AA}, excluding a narrow window around H$\gamma$: 4260--4430~{\AA}.
As described in Paper~\citetalias{Hu2020}, there appears to be a broad wing around {\hei}~$\lambda$5876 that poses challenges for accurately modeling.
Therefore our preferred {\hei} light curve used fluxes determined through direct integration rather than line profile fitting.
Figure~\ref{fig:fitexample} illustrates an example of the fitting applied to an individual-night spectrum, where the residuals relative to the spectral fluxes are very small ($\sim$2\%), indicating the reliability of our SFS.
The light curves of {\feii} and {\heii} emission lines are presented in Figure~\ref{fig:ccfmica34} and \ref{fig:ccfmica56} for each respective year.
Furthermore, we have investigated the potential of extending our fitting range to incorporate {\hei}~$\lambda$5876 and have contrasted the outcomes with those derived from direct integration.
Nevertheless, the {\hei} light curves from the SFS exhibit larger scatter compared to the integrated ones, which may be caused by the broad red wing of the line (see in Figure~\ref{fig:fitexample}).

In summary, we utilize the light curves of the 5100~{\AA} continuum, the broad {\hb} and {\hei} emission lines through the integration scheme, and the {\heii} and {\feii} lines from the SFS for subsequent analyses.
Table~\ref{tab:lc} lists parameters associated with the resultant light curves, while comprehensive representations of these light curves are plotted in Figure~\ref{fig:ccfmica34} and \ref{fig:ccfmica56}.
The 5100~{\AA} continuum light curves derived from the spectral data are strongly consistent with those obtained from the photometric data, thereby suggesting a high degree of accuracy in our spectral flux calibration.

\begin{deluxetable*}{cccccc}
\setlength{\tabcolsep}{18pt} 
\tablecolumns{6}
\tabletypesize{\footnotesize}
\tabcaption{Light Curves of Continuum and Several Emission Lines \label{tab:lc}}
\tablewidth{0pt}
\tablehead{
\colhead{JD - 2457000} & \colhead{\fcs} & \colhead{\fheii} 
& \colhead{\fhei} & \colhead{\fhb} & \colhead{\ffeii} 
}
\startdata
1616.672 & $5.494\pm0.064$ & $0.474\pm0.033$ & $0.569\pm0.002$ & $5.489\pm0.016$ & $4.739\pm0.046$ \\
1629.654 & $5.206\pm0.069$ & $0.614\pm0.035$ & $0.544\pm0.002$ & $5.437\pm0.019$ & $4.567\pm0.052$ \\
1638.644 & $5.084\pm0.049$ & $0.576\pm0.023$ & $0.561\pm0.002$ & $5.511\pm0.019$ & $4.526\pm0.035$ \\
1643.579 & $4.985\pm0.058$ & $0.550\pm0.029$ & $0.571\pm0.003$ & $5.617\pm0.023$ & $4.604\pm0.043$ \\
\enddata
\tablecomments{
The 5100~{\AA} continuum flux, {\fcs}, is expressed in units of 10$^{-15}$~{\ergscma}, whereas the units of all emission lines ({\fheii}, {\fhei}, {\fhb} and {\ffeii}) are 10$^{-13}$~{\ergscm}. \\
(This table is available in its entirety in a machine-readable form in the online journal.
A portion is shown here for guidance regarding its form and content.)
}
\end{deluxetable*}

\vspace{-8mm}
\section{Analysis} \label{sec:analysis}

\subsection{Variability} \label{subsec:fvar}
For each light curve determined in Section~\ref{sec:lc}, the intrinsic variability {\fvar} \citep{Rodriguez-Pascual1997} and its uncertainty {\sigfvar} \citep{Edelson2002} can be calculated as:
\begin{equation}
F_{\rm var}=\frac{(S^2-\Delta^2)^{1/2}}{\langle F \rangle}, \label{eq:fvar}
\end{equation}
and 
\begin{equation}
    \sigma_{F_{\rm var}}=\frac{1}{F_{\rm var}}\sqrt{\frac{1}{2N}}\frac{S^2}{{\langle F \rangle}^2}, \label{eq:err_fvar}
\end{equation}
where $\langle F \rangle$, $S^2$ and $\Delta^2$ represent mean flux, flux variance and mean square flux uncertainty, whose definitions are as follows: $\langle F \rangle=\frac{1}{N}\sum_{i=1}^{N}F_{i}$, $S^2=\frac{1}{N-1}\sum_{i=1}^{N}(F_{i}-\langle F \rangle)^2$ and $\Delta^2=\frac{1}{N}\sum_{i=1}^{N}\Delta_{i}^2$, where $\Delta_{i}$ represents the uncertainty of $F_{i}$.

\begin{figure}[!ht]
    \includegraphics[width=\columnwidth, trim=20 0 0 0]{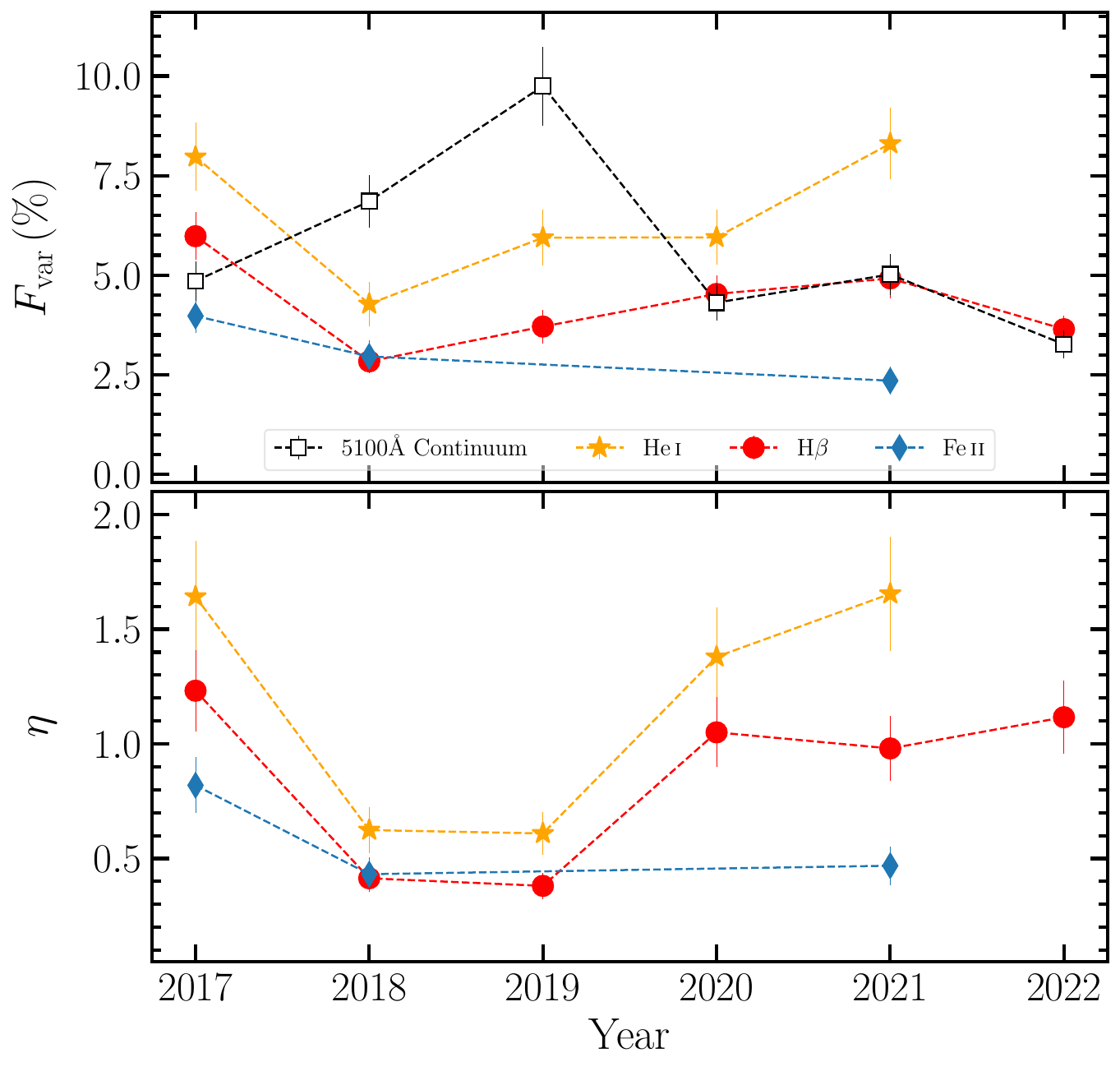}
    \caption{
    Top panel: The variability amplitudes of the 5100~{\AA} continuum and three emission lines.
    Bottom panel: The responsivity of three lines to the continuum, which are dimensionless.
    }
    \label{fig:fvareta}
\end{figure}

We also calculate the responsivity $\eta$ of the broad lines with respect to the 5100~{\AA} continuum, which is simply defined as the {\fvar} ratio of the line to the continuum \citep{Goad2014}:
\begin{equation}
    \eta=\frac{F_{\rm var, line}}{F_{\rm var, 5100}}, \label{eq:respon}
\end{equation}
Both $F_{\rm var}$ and $\eta$ are shown in Figure~\ref{fig:fvareta} and Table~\ref{tab:lc_char}.
The $F_{\rm var}$ values of {\hei}, {\hb} and {\feii} go up and down in a similar manner relative to each other through the years, while the 5100~{\AA} continuum in 2018 and 2019 displays large variability amplitudes.
Correspondingly, the $\eta$ in 2018 and 2019 are significantly lower compared with the other 4 years.
We quantitatively discuss the specifics in Section~\ref{subsec:reverberation}.

\begin{deluxetable*}{ccccllcllccccll}
\setlength{\tabcolsep}{6.6pt} 
\tabletypesize{\footnotesize}
\tablecaption{Light Curve Properties\label{tab:lc_char}}
\tablecolumns{15}
\tablehead{
\colhead{Year} & \multicolumn{2}{c}{5100~{\AA}}
&& \multicolumn{2}{c}{\heii} && \multicolumn{2}{c}{\hei}
&& \multicolumn{2}{c}{\hb} && \multicolumn{2}{c}{\feii} 
\\\cline{2-3}\cline{5-6}\cline{8-9}\cline{11-12}\cline{14-15}
& \colhead{$L_{\rm 5100}$} 
& \colhead{$F_{\rm var}$}
&& \colhead{$F_{\rm var}$} & \colhead{$\eta$}
&& \colhead{$F_{\rm var}$} & \colhead{$\eta$} 
&& \colhead{$F_{\rm var}$} & \colhead{$\eta$}
&& \colhead{$F_{\rm var}$} & \colhead{$\eta$}
\\
& \colhead{($10^{44}$~{\ergs})} 
& \colhead{(\%)} && \colhead{(\%)} &&& \colhead{(\%)} 
&&& \colhead{(\%)} &&& \colhead{(\%)} &
}
\startdata
2019 & $2.388\pm0.235$ & $9.7\pm1.0$ && $24.9\pm3.3$ & 2.6 && $5.9\pm0.7$ & 0.6 && $3.7\pm0.4$ & 0.4 && $3.1\pm0.5~\ast$ & 0.3 $\ast$ \\
2020 & $2.333\pm0.105$ & $4.3\pm0.4$ && $18.6\pm2.3$ & 4.3 && $6.0\pm0.7$ & 1.4 && $4.5\pm0.5$ & 1.1 && $5.0\pm0.5~\ast$ & 1.2 $\ast$ \\
2021 & $2.206\pm0.114$ & $5.0\pm0.5$ && $20.2\pm2.9$ & 4.0 && $8.3\pm0.9$ & 1.7 && $4.9\pm0.5$ & 1.0 && $2.4\pm0.3$ & 0.5 \\
2022 & $2.168\pm0.077$ & $3.3\pm0.3$ && $22.8\pm3.4~\ast$ & 7.0 $\ast$ && $4.1\pm0.5~\ast$ & 1.3 $\ast$ && $3.6\pm0.3$ & 1.1 && $3.8\pm0.4~\ast$ & 1.2 $\ast$ \\
\enddata
\tablecomments{
$L_{\rm 5100}$ represents the continuum luminosity at 5100~{\AA}.
Both $F_{\rm var}$ and $\eta$ are dimensionless.
The measurements annotated with $\ast$ have been excluded from the analysis for unreliable time lags.
}
\end{deluxetable*}

\vspace{-8mm}
\subsection{Time Lags Measurement} \label{subsec:lag}
One of the key aims of this study is to determine the reverberation lags of several broad emission lines.
We employ two distinct methods to conduct these measurements, which are briefly outlined as follows:

\emph{ICCF:} First, we employ the interpolated cross-correlation function (ICCF; \citealt{Gaskell1986, Gaskell1987, White1994}), a conventional technique in quasar RM, to determine the time delays.
The time lag we adopt is the centroid of the cross-correlation function (CCF), using only the part above 80\% of its peak value {\rmax} \citep{Koratkar1991, Peterson2004}. 
The uncertainties are estimated by the ``flux randomization/random subset sampling (FR/RSS)'' method, which uses the 15.87\% and 84.13\% quantiles of the cross-correlation centroid distribution (CCCD; \citealt{Maoz1989, Peterson1998, Peterson2004}).

\emph{MICA:} We also employ Multiple and Inhomogeneous Component Analysis (MICA\footnote{MICA is available at: \url{https://github.com/LiyrAstroph/MICA2}}; \citealt{Li2016, mica2_liyr}) to estimate the time delays, which is a Bayesian-based nonparameteric approach to constrain the transfer function from RM light curves.
MICA utilizes the damped random walk (DRW) model to reconstruct the light curves and capture the continuum variations, which expresses the transfer function as a sum of a series of relatively displaced Gaussians.
In this study, we simplify the parameterization of the transfer function by using a single or double Gaussian for each emission line.
Based on the quality of reconstructed light curves and Bayesian evidence calculated using the diffusive nested sampling algorithm~\citep{Brewer2009, Brewer2011} employed by MICA, we find that a double Gaussian is optimal for year 2022, while a single Gaussian is sufficient for the other three years (2019-2021).
Notably, in the MICA analysis of 2019, we detrended the light curves using a linear polynomial, resulting in a better fit for the reconstructed light curves compared to those without detrending.
Nevertheless, we stress that the time lag measurements were not significantly different between both them with ICCF.
Our MICA analysis allowed a high degree of flexibility: we did not impose any artificial constraints on the width of the transfer function, and also set sufficiently wide ranges of time lags, allowing negative responses.
The time lags are determined at the centroid of the Gaussians, while its associated lower and upper uncertainties are estimated as the 15.87\% and 84.13\% quantiles, respectively, of their posterior distributions generated by the Markov Chain Monte Carlo (MCMC) method.

Due to the contamination of emission lines in the $V$ band photometric flux, we prioritize the measured ICCF and MICA lags of {\fline} relative to the continuum {\fcs} given by spectroscopy as the final time lags.
All the reverberation results are listed in Table~\ref{tab:lags} and presented in Figures~\ref{fig:ccfmica34} and \ref{fig:ccfmica56}.
We deemed the measurements with absolute uncertainties exceeding 20 days unacceptable and excluded them.
Additionally, we also omited the time lag measurements that exhibited significant disparities (differing by more than 100\%) between ICCF and MICA.
We attribute these discrepancies to raw data quality issues that may have compromised the reliability of the light curves and time lag measurements.
Eventually, we use the results from ICCF for subsequent analyses and discussions.

\begin{deluxetable*}{ccrlccllccllccllccll}
\setlength{\tabcolsep}{4pt} 
\tabletypesize{\footnotesize}
\tablecaption{Reverberation Lags of Broad Lines measured by CCF and MICA \label{tab:lags}}
\tablecolumns{16}
\tablewidth{0.95\textwidth}
\tablehead{
\colhead{Year}
& \multicolumn{3}{c}{\heii} && \multicolumn{3}{c}{\hei}
&& \multicolumn{3}{c}{\hb} && \multicolumn{3}{c}{\feii}
\\\cline{2-4}\cline{6-8}\cline{10-12}\cline{14-16}
\colhead{}
& \colhead{\rmax} & \colhead{$\tau_{\rm ICCF}$} 
& \colhead{$\tau_{\rm MICA}$}
&& \colhead{\rmax} & \colhead{$\tau_{\rm ICCF}$} 
& \colhead{$\tau_{\rm MICA}$}
&& \colhead{\rmax} & \colhead{$\tau_{\rm ICCF}$} 
& \colhead{$\tau_{\rm MICA}$}
&& \colhead{\rmax} & \colhead{$\tau_{\rm ICCF}$} 
& \colhead{$\tau_{\rm MICA}$}
\\\colhead{}
& \colhead{} & \multicolumn{2}{c}{(day)}
&& \colhead{} & \multicolumn{2}{c}{(day)}
&& \colhead{} & \multicolumn{2}{c}{(day)}
&& \colhead{} & \multicolumn{2}{c}{(day)}
}
\startdata
2019 & 0.49 & $-4.2_{-3.9}^{+4.8}$ & $-5.3_{-1.3}^{+1.6}$ && 0.49 & $17.5_{-2.7}^{+5.1}$ & $12.6_{-2.5}^{+3.0}$ && 0.88 & $34.9_{-3.1}^{+3.1}$ & $29.5_{-2.8}^{+2.6}$ && 0.63 & $77.6_{-27.7}^{+16.3}$ & $63.0_{-26.7}^{+32.5}~\ast$ \\
2020 & 0.73 & $-4.6_{-5.5}^{+2.7}$ & $-4.0_{-2.6}^{+2.1}$ && 0.74 & $22.7_{-5.6}^{+13.4}$ & $32.0_{-5.4}^{+6.1}$ && 0.74 & $22.6_{-3.9}^{+6.6}$ & $25.1_{-3.9}^{+3.9}$ && 0.66 & $8.7_{-23.1}^{+15.7}$ & $45.6_{-46.9}^{+31.9}~\ast$ \\
2021 & 0.67 & $\ \ 3.6_{-5.0}^{+11.1}$ & $\ \ 5.8_{-4.4}^{+11.1}$ && 0.74 & $42.6_{-2.8}^{+11.7}$ & $53.2_{-3.7}^{+3.4}$ && 0.81 & $59.6_{-5.1}^{+5.1}$ & $58.8_{-3.5}^{+3.4}$ && 0.64 & $35.3_{-12.1}^{+14.3}$ & $41.3_{-6.1}^{+6.4}$ \\
2022 & 0.58 & $-2.4_{-5.5}^{+6.7}$ & $30.1_{-17.3}^{+11.8}~\ast$ && 0.48 & $25.3_{-0.6}^{+45.9}$ & $46.5_{-13.9}^{+13.8}~\ast$ && 0.66 & $55.1_{-18.1}^{+16.4}$ & $45.9_{-9.0}^{+7.1}$ && 0.68 & $93.7_{-23.3}^{+10.1}$ & $50.6_{-9.1}^{+3.6}~\ast$ \\
\enddata
\tablecomments{
The notation {\rmax} is the maximum correlation coefficient.
$\tau_{\rm MICA}$ was determined using a double-Gaussian transfer function for 2022 and a single Gaussian for the other years.
Table entries marked with $\ast$ have been excluded from the analysis.
}
\end{deluxetable*}

\begin{figure}[!ht]
    \includegraphics[width=\columnwidth, trim=20 0 0 0]{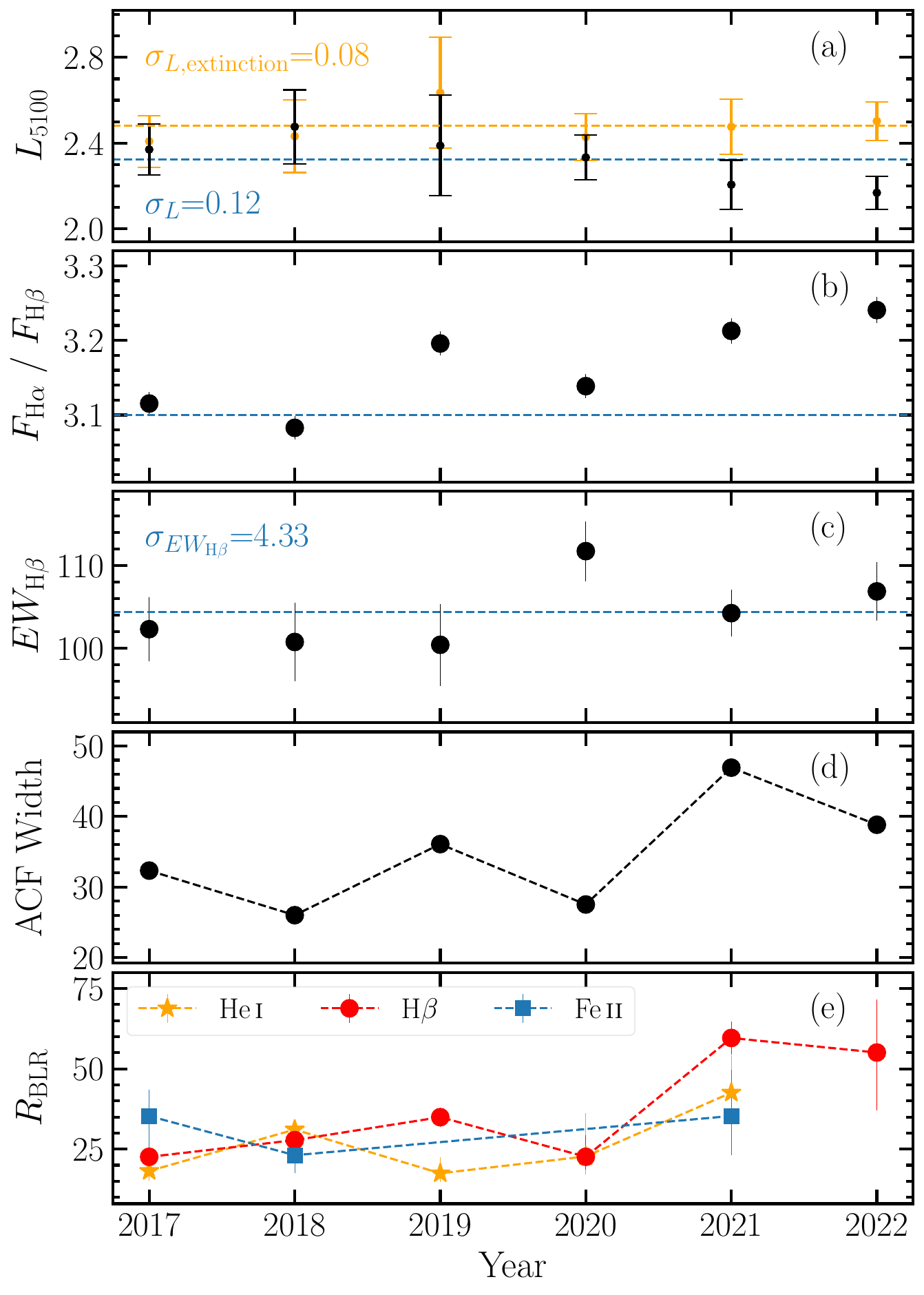}
    \caption{
    The evolution of several physical parameters over six years.
    From top to bottom:
    The mean luminosity at 5100~{\AA} (blue denotes observed values, while orange represents values corrected for dust extinction), Balmer decrement calculated by the flux ratio of {\ha} and {\hb}, $EW_{\rm H\beta}$, the widths of ACF and the BLR size.
    The luminosities are expressed in units of $10^{44}$~{\ergs};
    $EW_{\rm H\beta}$ are in units of {\AA};
    the ACF widths are in units of days, and the sizes of BLR are in units of light days.
    In panel (a) and (c), the mean values of all years are marked with dashed lines, and the standard deviations are displayed as well.
    $R_{\rm V}=$ 3.1 is marked with a blue dashed line in panel (b).
    In panel (e), black squares, red dots and blue triangles represent {\hb}, {\hei} and {\feii} emission lines, respectively.
    The data for 2017 and 2018 are from Paper~\citetalias{Hu2020}, while the data for 2019-2022 are from this work.
    Due to significant measurement uncertainties, the {\feii} lag for the years 2019, 2020 and 2022 as well as {\hei} lag for 2022 have all been excluded.
    }
    \label{fig:discuss}
\end{figure}

\vspace{-8mm}
An analysis of the results in panel (e) of Figure~\ref{fig:discuss} highlights intriguing yet perplexing phenomena regarding the BLR in {\pgpg}.
In line with the findings of Paper~\citetalias{Hu2020}, the {\hb} region exhibited slight changes during the initial two years.
The radii of the {\hei} and {\feii} regions however behaved differently.
In 2019, the radius of the {\hb} region increased by 50\% compared to the preceding two years.
In contrast, the {\hei} region contracted during that year.
In 2020, both {\hb} and {\hei} regions continued showing substantial changes compared to 2019, seemingly reverting to their original sizes observed in the first two years.
Note that, owing to significant measurement scatters, the results for {\feii} in 2019, 2020 and 2022, as well as {\hei} in 2022, are not completely reliable.
As a result these are not depicted in Figure~\ref{fig:discuss}.

Overall, the time lags of the {\hb} emission line exhibited a notable increase over six years, with a particularly significant jump observed in the last two years.
During the final two years, the {\hb} lags reached values that were two to three times larger than those observed in the preceding four years.
The measured time lags of {\heii} were slightly negative in most years (except the sixth year in which the results given by ICCF and MICA show large discrepancy). Considering the associated uncertainty, they were consistent with a zero lag throughout.

In terms of the relative magnitudes of the time delays observed for the three different emission lines, the BLR of {\pgpg} displayed a layered onion-like structure.
To be specific, in 2017 this ``onion'' was characterized by {\rfeii} $>$ {\rhb} $>$ {\rhei}, while this ordering reversed the following year, with {\rhei} $>$ {\rhb} $>$ {\rfeii} (refer to Section~5.1 in Paper~\citetalias{Hu2020}).
However, in the subsequent four years (2019-2022), we are unable to ascertain the stratified structure with the large uncertainties on the lag measurements for {\hei} and {\feii}.

\subsection{Line Width Measurements} \label{subsec:lineprofile}
For each observing season, we employ the line profile extracted from the mean or RMS spectrum to measure the width of the emission line.
The line width can be quantified using the full width at half maximum (FWHM) parameterization, as defined by \citet{Peterson2004}:
\begin{equation}
    {\rm FWHM}=\lambda_{2}-\lambda_{1},~
    P(\lambda_{i})=0.5~P_{\rm max}~(i=1, 2), \label{eq:fwhm}
\end{equation}
where $\lambda_{1}$ and $\lambda_{2}$ represent the wavelengths at which the line profile $P(\lambda)$ reaches half of its maximum value $P_{\rm max}$.
Alternatively, the line width can also be characterized by the line dispersion, which is determined as the difference between the mean square of the wavelength $\langle\lambda^2\rangle$ and the square of the central wavelength $\lambda_{0}$ \citep{Peterson2004}:
\begin{equation}
    \sigma_{\rm line}^2=\langle\lambda^2\rangle-\lambda_{0}^2,
\end{equation}
where
\begin{equation}
    \langle\lambda^2\rangle=\frac{\int\lambda^{2}~P(\lambda)~d\lambda}{\int~P(\lambda)~d\lambda},~
    \lambda_{0}=\frac{\int\lambda~P(\lambda)~d\lambda}{\int~P(\lambda)~d\lambda}.
\end{equation}

The fitting procedure for {\hb}, {\hei} and {\feii} was described in Section~\ref{subsec:fit}.
To account for uncertainties from RMS spectra, we employed a Monte Carlo (MC) simulation approach \citep{Chen2023}.
Firstly, we generated $N$ mean and RMS spectra through bootstrap sample selection.
This involved iteratively constructing new mean and RMS spectra by randomly selecting a subset of $m-n$ spectra using Equation~\ref{eq:mean} and ~\ref{eq:rms}, where $m$ denotes the number of spectra selected in each iteration and $n$ represents the number of duplicated spectra.
Subsequently, we performed a series of steps including applying SFS to the obtained $N$ mean and RMS spectra, extracting line profiles, and calculating line widths, to obtain the distributions of line widths.
The standard deviations of these distributions were determined to quantify the scatters of the line widths.
For this investigation we performed $N=$ 1000 Monte Carlo simulations.

For each line profile (including those obtained from the MC simulations), the line width is corrected by comparing the width of {\oiii} in the corresponding spectrum to those from previous high-resolution observations. 
We adopt a value of 350 {\kms} (obtained from \citet{Whittle1992}) as a reference FWHM of {\oiii}~$\lambda$5007 \citep{Grier2012, Hu2020, Chen2023}.
We use this value to correct for the instrumental spectral broadening.
The resulting measurements of line widths are all presented in Table~\ref{tab:lw_lines}.

\begin{deluxetable*}{cllclccl}
\setlength{\tabcolsep}{9.8pt} 
\tabletypesize{\footnotesize}
\tablecaption{Velocity Widths of Several Broad Emission Lines and Black Hole Masses\label{tab:lw_lines}}
\tablecolumns{8}
\tablewidth{0.9\textwidth}
\tablehead{
\colhead{Year}
& \colhead{\heii} & \colhead{\hei}
& \multicolumn{4}{c}{\hb} & \colhead{\feii}
\\\cline{4-7}
& \colhead{FWHM} & \colhead{FWHM}
& \colhead{FWHM} & \colhead{\rhb}
& \colhead{$M_{\rm BH} (f=0.5)$} 
& \colhead{$\lg{\dot{\mathscr{M}}}$}
& \colhead{FWHM}
\\
& \colhead{(\kms)} & \colhead{(\kms)} 
& \colhead{(\kms)} & \colhead{(lt-days)}
& \colhead{($\times10^{7}M_{\odot}$)} 
&& \colhead{(\kms)}
}
\startdata
2019 & $7487.8\pm1124.8$ & $2451.4\pm18.9$ & $2081.7\pm45.5$ & $35.0_{-3.1}^{+3.1}$ & $1.48_{-0.15}^{+0.14}$ & 
$1.72_{-0.12}^{+0.10}$ & $1815.3\pm105.1$ \\
2020 & $9594.5\pm1033.0$ & $2347.2\pm9.6$ & $1976.2\pm37.0$ & $22.6_{-3.9}^{+6.6}$ & $0.86_{-0.15}^{+0.25}$ & 
$2.17_{-0.20}^{+0.20}$ & $1698.5\pm72.4$ \\
2021 & $9352.6\pm1360.3$ & $2461.2\pm14.4$ & $2021.0\pm68.9$ & $59.6_{-5.1}^{+5.1}$ & $2.38_{-0.26}^{+0.26}$ & 
$1.25_{-0.12}^{+0.09}$ & $1704.0\pm96.1$ \\
2022 & $7134.0\pm896.1$ & $2430.7\pm8.9$ & $1944.5\pm54.8$ & $55.1_{-18.1}^{+16.4}$ & $2.03_{-0.68}^{+0.62}$ & 
$1.38_{-0.48}^{+0.21}$ & $1677.9\pm64.1$ \\
\enddata
\tablecomments{
All the line widths listed here are FWHM measured from the mean spectra.
Note that the {\heii} FWHM are fixed to its values of RMS spectra.
$M_{\rm BH}$ are calculated using the virial factor associated with the FWHM of the mean spectra \citep{Ho2014}.
{\rhb} are from the ICCF measurements.
}
\end{deluxetable*}

\begin{deluxetable}{cllcc}
\setlength{\tabcolsep}{4pt} 
\tabletypesize{\footnotesize}
\tablecaption{$\sigma_{\rm line}$ of {\hb} RMS spectra and Black Hole Masses \label{tab:lw_hb}}
\tablecolumns{5}
\tablewidth{0.95\textwidth}
\tablehead{
\colhead{Year} & \colhead{$\sigma_{\rm line}$(RMS)}
& \colhead{\rhb}
& \colhead{$M_{\rm BH} (f=3.2)$} & \colhead{$\lg{\dot{\mathscr{M}}}$} \\
& \colhead{(\kms)} & \colhead{(lt-days)} & \colhead{($\times10^{7}M_{\odot}$)} &
}
\startdata
2019 & $1382.5\pm202.8$ & $35.0_{-3.1}^{+3.1}$ & $4.17_{-1.28}^{+1.28}$ & $0.82_{-0.43}^{+0.21}$ \\
2020 & $890.4\pm44.0$ & $22.6_{-3.9}^{+6.6}$ & $1.12_{-0.22}^{+0.35}$ & $1.94_{-0.23}^{+0.21}$ \\
2021 & $1106.9\pm99.0$ & $59.6_{-5.1}^{+5.1}$ & $4.56_{-0.91}^{+0.91}$ & $0.69_{-0.23}^{+0.15}$ \\
2022 & $722.0\pm104.1$ & $55.1_{-18.1}^{+16.4}$ & $1.79_{-0.78}^{+0.74}$ & $1.49_{-0.90}^{+0.26}$ \\
\enddata
\tablecomments{
$M_{\rm BH}$ calculated with the virial factor obtained from $\sigma_{\rm line}$ for the RMS spectra \citep{Ho2014}.
{\rhb} are from the ICCF measurements.
}
\end{deluxetable}

\vspace{-8mm}
\subsection{Velocity-resolved Delay} \label{subsec:vrd}
In order to investigate the geometry and kinematics of the BLR in {\pgpg} and any potential changes, we have calculated the velocity-resolved time delays of the broad {\hb} line \citep[e.g., ][]{Bentz2009a, Grier2013a, Du2016, Xiao2018a, Zhang2019, Hu2020a, Feng2021, Feng2021a, Li2021, Kollatschny2022, Bao2022, Li2022}.
Initially, we divide the {\hb} emission-line profile into multiple velocity bins, ensuring that the integral fluxes of the corresponding RMS spectrum are equal.
This allows us to obtain light curves for each velocity bin and measure the time delays relative to the 5100~{\AA} continuum flux {\fcs} by using CCF, as detailed in Appendix~\ref{app:vrd}.

\begin{figure*}[!ht]
    \includegraphics[width=2\columnwidth, trim=0 0 40 0]{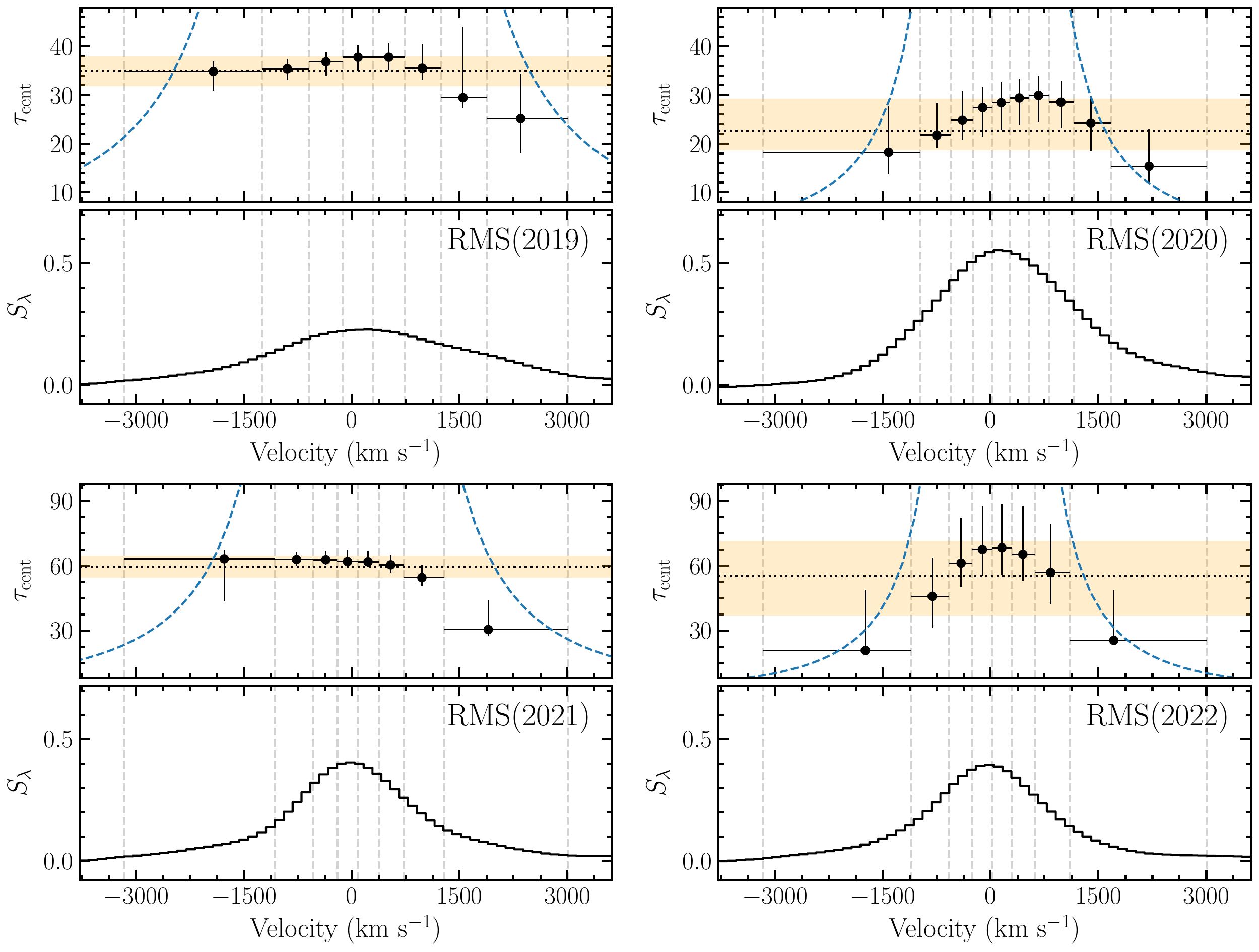}
    \caption{
    Velocity-resolved time delay measurements of the {\hb} profiles in the RMS spectra for four years.
    The upper-panels show the lags (black points with error bars) of fluxes determined using different velocity bins (separated with grey dashed lines) relative to {\fcs}.
    The orange region shows the {\hb} lags determined by comparing {\fhb} and {\fcs} with the CCF in Section~\ref{subsec:lag}.
    The blue dashed curves show the virial envelope for a Keplerian disk with an inclination of $\cos{i}=$ 0.75 and the estimated black hole mass from Table~\ref{tab:lw_hb}.
    The RMS spectrum (with continuum subtracted) is presented in the lower panel.
    The lag, $\tau_{\rm cent}$ is in units of days, while
    the flux of RMS spectra $S_{\rm \lambda}$ is expressed in units of 10$^{-15}$~{\ergscma}.
    }
    \label{fig:vrd}
\end{figure*}

The velocity-resolved time delays and the RMS profiles for the four years are presented in Figure~\ref{fig:vrd}.
As demonstrated in Figure~6 of Paper~\citetalias{Hu2020} and the top two panels of our Figure~\ref{fig:vrd}, the BLR in 2017 exhibits the signature of a Keplerian/virialized disk, while the system appears to transition towards an inflow structure in 2018.
Subsequently, for the following four years, the BLR geometry is still changing, displaying a pattern of inflow-virial-inflow-virial.

\section{Discussion} \label{sec:discussion}
\subsection{The R-L Relation} \label{subsec:rl}
\begin{figure}[!ht]
    \includegraphics[width=\columnwidth]{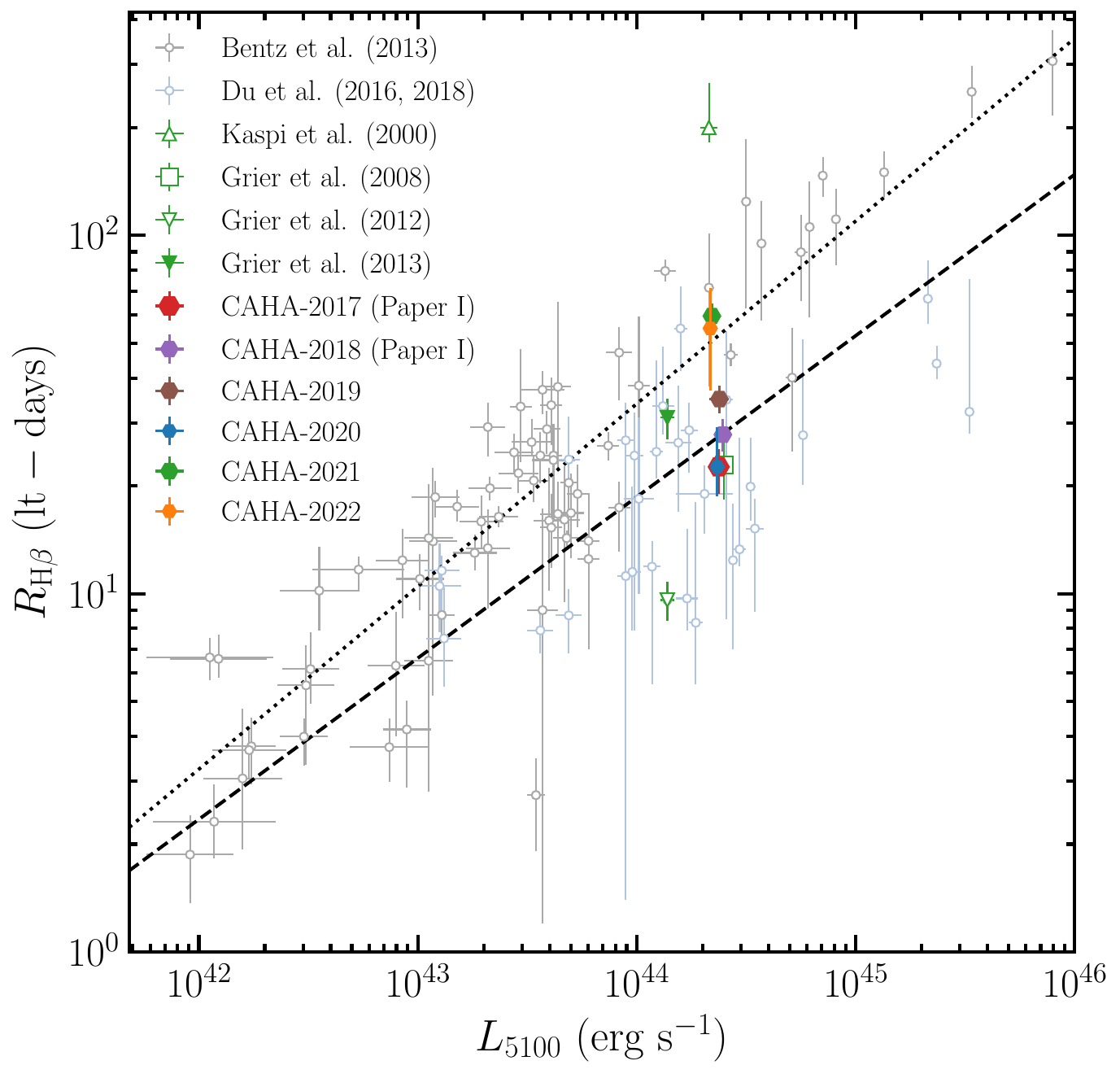}
    \caption{
    {\rl} relationship of {\pgpg} determined for each of six CAHA seasons (Paper~\citetalias{Hu2020} and this work) and previous campaigns \citep{Kaspi2000, Grier2008, Grier2012, Grier2013a}.
    The background samples from \citet{Bentz2013} (gray) and \citet{Du2016, Du2018} (light steel blue) are plotted for comparison.
    The dotted and dashed lines define this relationship for the RM samples of \citet{Du2018} with {\mathmdot} $<$ 3 and {\mathmdot} $\geqslant$ 3, respectively.
    }
    \label{fig:rl}
\end{figure}

The intrinsic {\rl} relationship for {\pgpg} from our SEAMBH-CAHA campaign, along with previous results from \citet{Kaspi2000, Grier2008, Grier2012} are presented in Figure~\ref{fig:rl}.
For comparison, the samples from \citet{Bentz2013} and \citet{Du2016, Du2018} are added as background.
The data point obtained from the campaign conducted by \citet{Kaspi2000} appears to be a clear outlier in the {\rl}, which is likely due to poor sampling \citep{Grier2008}.
Similarly, the data point from \citet{Grier2012} appears to deviate from the relationship established by \citet{Bentz2013}, but it aligns with the relationship proposed by \citet{Du2018}.
Notably, \citet{Grier2013a} and \citet{Bentz2013} argued that the time lag derived from \citet{Grier2012} should be 31$\pm$4 days when correcting for the absence of crucial sampling epochs.
If this correction is applied, the data point will coincide with the samples from \citet{Bentz2013}.
Another point of {\pgpg} is from \citet{Grier2008}.
Despite the relatively short duration of the campaign, the observed time delay for {\pgpg} remains consistent with the time delays observed in other samples, including the data from the current study.
However, these previous observations have either low cadence or short duration, which could bias the time-lag measurements.

The six-year CAHA observations, comprising two years reported in Paper~\citetalias{Hu2020} and four presented in this work, are characterised by uniform high-cadence sampling, and are thus unbiased, allowing us to study the behavior of the time lag over long time spans.
Our results reveal a significant factor of $\sim$2 variation in the size of {\hb} BLR despite nearly constant luminosity ($\sigma_{L}\sim$ 0.12$\times10^{44}$~{\ergs}, see black points in panel (a) of Figure~\ref{fig:discuss}).
It is worth noting that~\citet{Zastrocky2024} also found some other objects exhibiting changing time lags without corresponding luminosity changes.
This kind of discrepancy contradicts the linear relationship between the BLR radius and the continuum luminosity established by the collective data from other RM campaigns \citep{Bentz2013, Du2016, Du2018}.

\begin{figure*}[!ht]
    \includegraphics[width=2\columnwidth, trim=0 0 40 0]{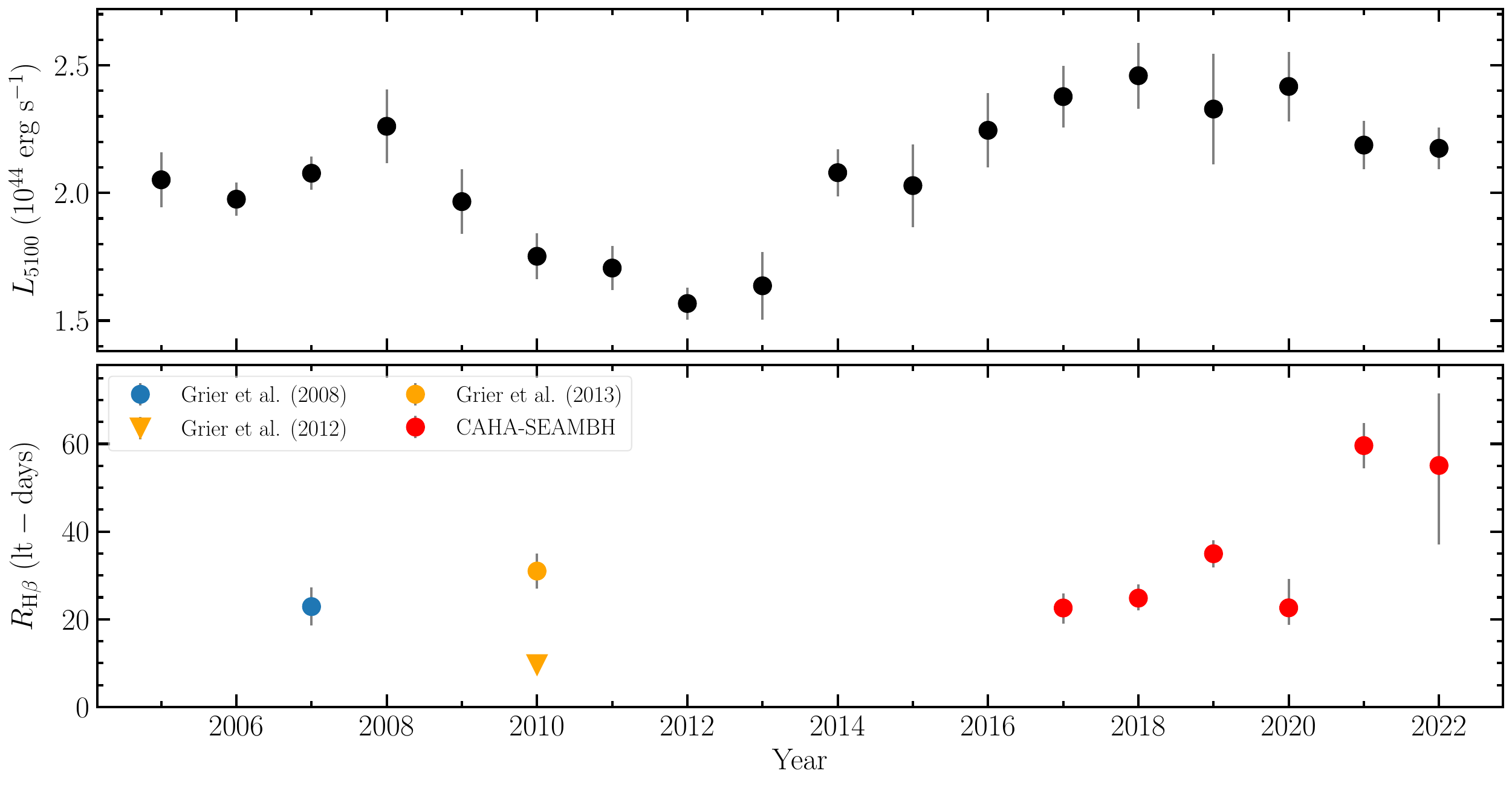}
    \caption{
    The variations of the optical luminosity ({\lcs}) and the size of the broad {\hb} region ({\rhb}) from 2005-2022.
    In the lower panel, the blue and orange points are from \citet{Grier2008, Grier2012, Grier2013a}, while the CAHA-SEAMBH points are plotted in red.
    }
    \label{fig:llag}
\end{figure*}

In addition, we have compiled the archival photometry data, including: the Catalina Real-Time Transient Survey (CRTS\footnote{\url{http://crts.caltech.edu/}}; \citealt{Drake2009}), the All-Sky Automated Survey for SuperNovae (ASAS-SN\footnote{\url{http://www.astronomy.ohio-state.edu/asassn/index.shtml}}; \citealt{Shappee2014, Kochanek2017}) and the Zwicky Transient Facility (ZTF\footnote{\url{https://www.ztf.caltech.edu}}; \citealt{Bellm2019a}).
We then intercalibrate these photometry datasets with our CAHA 5100~{\AA} continuum data by applying a Bayesian-based package PyCALI\footnote{PyCALI is available at: \url{https://github.com/LiyrAstroph/PyCALI}} \citep{Li2014, pycali_liyr}, which describes the AGN variability using the DRW model \citep[e.g., ][]{Kelly2009, Kelly2011, Zu2011, Li2013, Li2014, Li2018, Lu2019a}).
The detailed light curves are shown in Appendix~\ref{app:longtermlc} and the mean luminosity has been calculated for each year.
The luminosities at 5100~{\AA} ({\lcs}) and the size of the broad {\hb} region ({\rhb}) for all seasons are shown in Figure~\ref{fig:llag}.
In terms of the amplitude of the long-term light curves, the size of the BLR has increased by $\sim$100\% ($\sim$112\% for {\hb} emitting region and $\sim$90\% for {\hei} region), while the luminosity changed less.
Looking at the variation in the 5100~{\AA} continuum luminosity, its maximum increase (from 2012 to 2018) is approximately 70\%, which is lower than that of the {\rhb}.
The long-term evolving patterns of the continuum and broad {\hb} time delay are however inconsistent with each other.
The continuum shows a slight decrease during the middle years, whereas the {\hb} time delay initially exhibits shorter values and longer ones subsequently.
Consequently, the alterations in the dynamics of the BLR driven by radiation pressure, in, e.g., NGC~5548 \citep{Lu2016, Lu2022} and NGC~4151 \citep{Chen2023}, cannot account for the variations observed in {\pgpg}.
In subsequent discussions, we find that the time lag variations are not due to physical effects but are more likely caused by measurement biases arising from the ``geometric dilution'', see details in Section~\ref{subsec:otherfactors}.

\subsection{Black-hole Mass and Accretion Rate} \label{subsec:mbh}
The black hole mass $M_{\bullet}$ can be determined by the virial relation:
\begin{equation}
    M_{\bullet}=f\times\frac{c~\tau_{\rm BLR}~V^2}{G}\equiv~f\times{\rm VP}, \label{eq:VP}
\end{equation}
where $c$ and $G$ represent the speed of light and gravitational constant, respectively.
$\tau_{\rm BLR}$ is the time lag measured in Section~\ref{subsec:lag}.
$V$ represents the line width of broad emission line, which is determined by either FWHM or velocity dispersion $\sigma_{\rm line}$ in Section~\ref{subsec:lineprofile}.
$f$ is the virial factor, which can be scaled by assuming that AGNs follow the same {\mbh} $- \sigma_{\ast}$ relation as inactive galaxies \citep[e.g.,][]{Onken2004, Woo2010, Woo2015, Graham2011}.
According to the classification scheme proposed by \citet{Ho2014}, {\pgpg} is categorized as an AGN with a pseudobulge, which typically exhibits smaller (about half) $f$ compared to classical bulges and ellipticals.
Furthermore, it is preferable to utilize different values of $f$ when employing different line width measurements \citep{Ho2014}.
For instance, a value of $f=$ 0.5 can be adopted when estimating the black hole mass by using FWHM of the mean spectra as adopted in Paper~\citetalias{Hu2020}.

According to the standard accretion disk model proposed by \citet{Shakura1973}, the dimensionless accretion rate (Eddington ratio) is defined as: $\dot{\mathscr{M}}=\dot{M_{\bullet}}/L_{\rm Edd}~c^{-2}$, where $\dot{M_{\bullet}}$ represents the accretion rate and $L_{\rm Edd}$ denotes the Eddington luminosity.
We can also estimate it with the formula \citep{Du2016a}:
\begin{equation}
    \dot{\mathscr{M}}=20.1\left(\frac{\ell_{44}}{\cos{i}}\right)^{3/2}M_{7}^{-2}, \label{eq:mdot}
\end{equation}
where $M_7=M_{\bullet}/10^7~M_{\odot}$ and $\ell_{44}=L_{\rm 5100}/10^{44}$~{\ergs} (with $L_{\rm 5100}$ representing the luminosity at 5100~{\AA}).
The parameter $i$ represents the inclination angle between the accretion disk and the line of sight.
Here we adopt a value of $\cos{i}=$ 0.75, which corresponds to a mean disk inclination for type 1 AGNs \citep{Du2016a, Hu2020, Li2021}.
The estimated black hole masses and accretion rates are presented in Table~\ref{tab:lw_lines}.

By combining the $M_{\bullet}-\sigma_{\ast}$ relation \citep{Ho2015} with the bulge velocity dispersion of $\sigma_{\ast}=163\pm19$ {\kms} \citep{Grier2013a}, the upper limit black hole mass of {\pgpg} becomes $6.5\times10^{7}M_{\odot}$.
Additionally, as mentioned in Paper~\citetalias{Hu2020}, the velocity-resolved delay measurements and MCMC analysis from \citet{Grier2017a} indicate a conservative estimate for the black hole mass of $10^{6.87_{-0.23}^{+0.24}}M_{\odot}$, i.e. $7.41_{-3.05}^{+5.47}\times10^{6}M_{\odot}$, determined using the FWHM of the mean {\hb} spectrum in 2017.
However, most RM studies consider $\sigma_{\rm line}$ of the {\hb} RMS spectra a more reliable measurement of black hole mass \citep[e.g., ][]{Peterson2004, Hu2020a, Bonta2020, Bao2022}.
Therefore, we also present the results calculated using $\sigma_{\rm line}$ of the {\hb} RMS spectra in Table~\ref{tab:lw_hb}, which exceed the values derived from the mean FWHM and also exhibit larger systematic uncertainties.
These discrepancies between the measurements obtained using different line widths also indicate significant systematic errors in the estimation of black hole mass.
We propose that simply using the FWHM or $\sigma_{\rm line}$ calculated from line profiles may not perfectly track the dynamics of the BLR gas.
Therefore, employing BLR dynamical modeling \citep[e.g., ][]{Li2013, Li2018} in future investigations is essential to gain more insight into evolution of BLR geometry and kinematics, and to provide more reliable estimates of the black hole mass.

\subsection{Reverberation} \label{subsec:reverberation}
We calculated the traditional variability amplitudes ({\fvar}) of the 5100~{\AA} continuum and emission lines, and quantified the responsivities ($\eta$) of the emission lines as the ratio of the emission line's {\fvar} to the continuum's (see Section~\ref{subsec:fvar}).
The values of {\fvar} and $\eta$ for each year are illustrated in Figure~\ref{fig:fvareta}, which excludes data points that were deemed unreliable as a result of inconsistencies in the light curve and time delay measurements (refer to Section~\ref{subsec:lag}).
We now discuss if the dramatic variations in the {\hb} time lag measurements for {\pgpg} could be related to changes in the line response.

As illustrated in the Figure~\ref{fig:fvareta}, the responsivity of {\hb} in the years 2018 and 2019 (0.41 and 0.38) are significantly lower than those in the other four years ($\sim$1.09 on average), indicating a weaker response of the {\hb} emission line to the continuum's drastic variability during these two years.
However, during these two years, the variability amplitude of {\hb} did not significantly differ from other years and the {\hb} still displayed clear variability characteristics following the continuum.
Therefore, despite the weaker response of {\hb}, the time delay measurements still remain reliable.
Even excluding these two years of weak response, the time lag measurements in the last two years increased by 2-3 times compared to 2017 and 2020, despite similar $\eta$.
Therefore, the significant changes in the {\hb} time lag measurement cannot be related to the line response and evidently require a comprehensive explanation.

\subsection{Other Factors} \label{subsec:otherfactors}
In this section, we discuss other factors that may affect the measurements of the ionizing continuum flux and the time lag, including dust extinction, the equivalent width of {\hb} emission line ($EW_{\rm H\beta}$, as an indicator of SED) and the variability timescale of the continuum.

Usually, the broad-line Balmer decrement estimated by the flux ratio of broad {\ha} to {\hb} can be an indicator of the extinction of AGNs
\citep{Netzer1975, Dong2007, Lu2019b, Li2022c, Ma2023a}.
Thus, we also calculate the mean Balmer decrement of {\pgpg} for each year directly from the mean spectra shown in panel (b) of Figure~\ref{fig:discuss}.
The fluxes of {\ha} and {\hb} are given by direct integration, and the contamination from the narrow lines are negligible given their weak fluxes.
The Balmer decrement values are directly obtained by comparing the fluxes of {\ha} and {\hb} year by year.
For each year, we employ the Balmer decrement to evaluate the dust extinction, and subsequently apply extinction corrections to the continuum \citep{Cardelli1989}.
The corrected continuum luminosities still exhibit consistent behavior over the entire six year span with a smaller dispersion ($\sim$0.08$\times10^{44}$~{\ergs}) than before ($\sim$0.12$\times10^{44}$~{\ergs}), as depicted by the orange data in panel (a) of Figure~\ref{fig:discuss}.
The mean correction for extinctions of the continuum luminosity over six years is $\sim$7\%, while the maximum correction is only $\sim$15\%.
Hence, we eliminate the impact of dust extinction on the continuum luminosity as the reason for the measured deviations in the {\rl} relationship.

It can be seen from Figure~\ref{fig:mr} that the mean spectra across the six-year period are almost the same, indicating a corresponding uniformity in the equivalent widths of all emission lines.
As a measure of the efficiency by which ionizing continuum photons are converted into line photons \citep{Korista2004}, we determined $EW_{\rm H\beta}$ effectively the ratio of {\hb} and 5100~{\AA} continuum, i.e. $EW_{\rm H\beta}=F_{\rm H\beta}/F_{\rm 5100}$, and the results are shown in panel (c) of Figure~\ref{fig:discuss}.
Our analysis only revealed a rather small variation in the average $EW_{\rm H\beta}$ across the six years, with a variation of only $\sim$4.15\% (where $\sigma_{EW_{\rm H\beta}}=$ 4.33~{\AA} and $\overline{EW_{\rm H\beta}}=$ 104.41~{\AA}).
Consequently, it can be inferred that the doubling of the measured time lag cannot be caused by the variation in the SED, which would have an effect on the equivalent widths of emission lines.

In Paper~\citetalias{Hu2020}, the continuum variability timescales of the first two years are compared with the widths of the ACFs.
The width of the ACF is defined as the width at half maximum, i.e., the width at the point where the correlation coefficient is 0.5.
We calculated the widths of ACFs for all six years and show their evolution with time in panel (d) of Figure~\ref{fig:discuss}.
In 2019, the continuum light curve shows a sharp upward trend (see Figure~\ref{fig:ccfmica34} and {\fvar} = 9.75), which may enlarge the measurement of the continuum variability timescale in that year.
Hence, in the computation of the ACF width for 2019, we initially applied a linear detrending to the continuum light curve \citep[e.g., ][]{Welsh1999, Denney2010, Grier2012a, Dietrich2012, Peterson2014a, Pei2017, Fausnaugh2017, Zhang2019}.
The mean width of ACF remained at $\sim$30.5 days in the first four years, and showed an upward trend in the last two years, increasing by $\sim$40\% (46.9 days and 38.8 days respectively).
This is consistent with the phenomenon of increasing measured time delays, i.e., there is a positive correlation between the continuum variability timescale and {\hb} time lag.

Based on the definitions of ACF and CCF, we can derive that the CCF is the convolution of the transfer function and the continuum ACF~\citep{Peterson1997}:
\begin{equation}
    F_{\rm CCF}(\tau) = \int_{-\infty}^{+\infty}F_{\rm ACF}(\tau-\tau^\prime) \Psi(\tau^\prime) \difd\tau^\prime,
    \label{eq:acf}
\end{equation}
where $\Psi$ is the transfer function.
Therefore, the ACF representing continuum variability timescale directly affects the CCF and measured time lag, with their relationship being nonlinear.
As indicated by detailed simulations in~\citet{Goad2014}, when the continuum variability timescale is faster than the time delay corresponding to the furthest extent of the BLR gas, the emission-line responses may be partially smeared (see their Figure~9), which is known as ``geometric dilution''~\citep{Robinson1990, Pogge1992, Gilbert2003} of the BLR.
For {\pgpg}, during the initial years, the continuum from the central ionization source changed swiftly, causing the emission line response from the outer part of the BLR to be smeared due to ``geometric dilution''.
Therefore, only the response from the inner part is detected, resulting in a smaller measured lag than the full extent of the BLR.
In contrast, during the latter two years, the damped continuum variability timescale gradually becomes larger, allowing the outer region of the BLR to respond and yield longer time lags.

Based on this interpretation, we suggest that the size of the BLR gas in {\pgpg} may extend beyond 60 light days.
During periods of rapid continuum fluctuations, only the inner part of the BLR exhibits a strong response and the measured short time lags are actually responsivity-weighted, capturing the reverberation concentrated at smaller radii rather than accurately reflecting the entire BLR size.
Consequently, in the case of small luminosity variations, using this short responsivity-weighted time lag may cause significant scatter in the intrinsic {\rl} relation of the AGN.

\section{Summary} \label{sec:summary}
We have continued the RM monitoring campaign in the CAHA-SEAMBH project for {\pgpg} after Paper~\citetalias{Hu2020} and analyzed the reverberations of several broad emission lines in all six years (2017-2022).
We find the following:
\begin{itemize}
\item
The reverberation lags of the broad emission lines, especically {\hb}, demonstrate a conspicuous increase over a span of six observational seasons, with notably substantial increases observed during the final two years, culminating in values of $\sim$60 days, which are $\sim$2 times larger than those observed in the preceding four years.
\item
The luminosity variation in the 5100~{\AA} continuum over six years is rather small, with a standard deviation of approximately 10\%.
Meanwhile, the lag of {\hb} changes by a factor of two.
The behavior of {\pgpg} is inconsistent with the typical {\rl} relationship of most sources studied so far with RM.
\item
After comprehensive consideration of various physical parameters (including the damped continuum variability timescale and amplitudes, the responsivity of the emission line to the continuum, the equivalent width of the emission line, and Balmer decrement), we tentatively propose that the changes of the BLR size measurement may be affected by the effect of ``geometric dilution''.
In light of this interpretation, the size of the BLR of {\pgpg} may extend to $\sim$60 days, while during periods of rapid continuum fluctuations, the measured short responsivity-weighted time delays primarily capture the reverberation concentrated in the inner part of the BLR.
\item
The velocity-resolved time delays of the broad {\hb} emission exhibit distinct characteristics across different observational seasons, transitioning cyclically between manifestations of virial motion and inflow, with the following sequence exhibited during the six years studied: virial-inflow-inflow-virial-inflow-virial.
BLR dynamical modeling with the {\hb} RM data will be analyzed in a forthcoming paper to gain more insight into the evolution of the BLR geometry and kinematics.
\end{itemize}

\section*{acknowledgements}
We acknowledge the support of the staff of the CAHA 2.2m telescope.
This work is based on observations collected at the Centro Astron\'omico Hispanoen Andaluc\'ia (CAHA) at Calar Alto, operated jointly by the Andalusian Universities and the Instituto de Astrof\'isica de Andaluc\'ia (CSIC).
We acknowledge financial support from the National Key R\&D Program of China (2021YFA1600404), the National Natural Science Foundation of China (NSFC; 11833008, 11991050, and 12333003).
Y.R.L. acknowledges financial support from the NSFC through grant No. 12273041 and from the Youth Innovation Promotion Association CAS.
C.H. acknowledges financial support from NSFC grants NSFC-12122305.
P.D. acknowledges ﬁnancial support from NSFC grants NSFC-12022301 and 11991051.
L.C.H. acknowledges financial support from the NSFC (11721303, 11991052, 12011540375, and 12233001), the National Key R\&D Program of China (2022YFF0503401), and the China Manned Space Project (CMS-CSST-2021-A04, CMS-CSST-2021-A06).

We acknowledge the efforts for public data from CTRS, ASAS-SN and ZTF.
The Catalina Sky Survey is funded by the National Aeronautics and Space Administration under Grant No. NNG05GF22G issued through the Science Mission Directorate Near-Earth Objects Observations Program.
The CRTS survey is supported by the US National Science Foundation under grants AST-0909182 and AST-1313422.
ASAS-SN is supported by the Gordon and Betty Moore Foundation through grant GBMF5490 to the Ohio State University and NSF grant AST-1515927.
Development of ASAS-SN has been supported by NSF grant AST-0908816, the Mt. Cuba Astronomical Foundation, the Center for Cosmology and Astro-Particle Physics at the Ohio State University, the Chinese Academy of Sciences South America Center for Astronomy (CASSACA), the Villum Foundation, and George Skestos.
ZTF is supported by the National Science Foundation under Grant No. AST-2034437 and a collaboration including Caltech, IPAC, the Weizmann Institute for Science, the Oskar Klein Center at Stockholm University, the University of Maryland, Deutsches Elektronen-Synchrotron and Humboldt University, the TANGO Consortium of Taiwan, the University of Wisconsin at Milwaukee, Trinity College Dublin, Lawrence Livermore National Laboratories, and IN2P3, France.
Operations are conducted by COO, IPAC, and UW.

\facilities{CAHA 2.2m}

\software{IRAF~\citep{Tody1986, Tody1993}, DASpec~\citep{daspec_dupu}, MICA~\citep{Li2016, mica2_liyr}, PyCALI~\citep{Li2014, pycali_liyr}.}

\appendix

\section{CCF analysis in each velocity bin} \label{app:vrd}
\begin{figure*}[!ht]
	\includegraphics[width=2\columnwidth, trim=0 0 60 0]{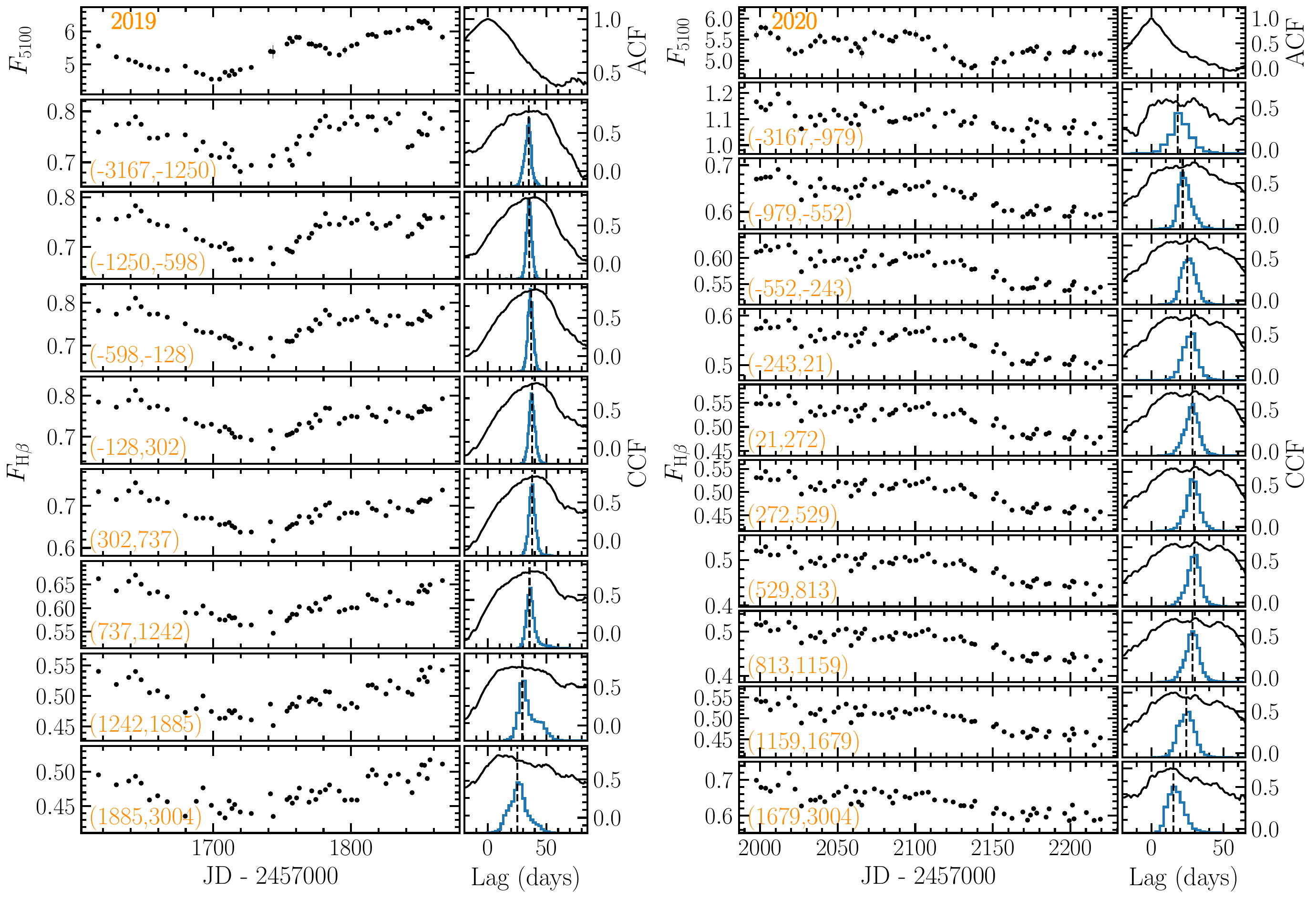}
    \caption{
    CCF analysis in each velocity bin with equal flux in the RMS spectrum in 2019 (left) and 2020 (right).
    The topmost panels show the 5100~{\AA} continuum light curves and ACFs.
    The rest of the panels show the {\hb} light curves (black points with error bars), ICCFs (black lines), CCCDs (blue histograms) and $\tau_{\rm cent}$ (black dashed lines) in each velocity bin (orange pairs of numbers with brackets).
    The units of the {\fcs} and {\hb} light curves are 10$^{-15}$~{\ergscma} and 10$^{-13}$~{\ergscm}, respectively.
    The units of velocity bin edges are {\kms}.}
    \label{fig:bin34}
\end{figure*}

\begin{figure*}[!ht]
    \includegraphics[width=2\columnwidth, trim=0 0 60 0]{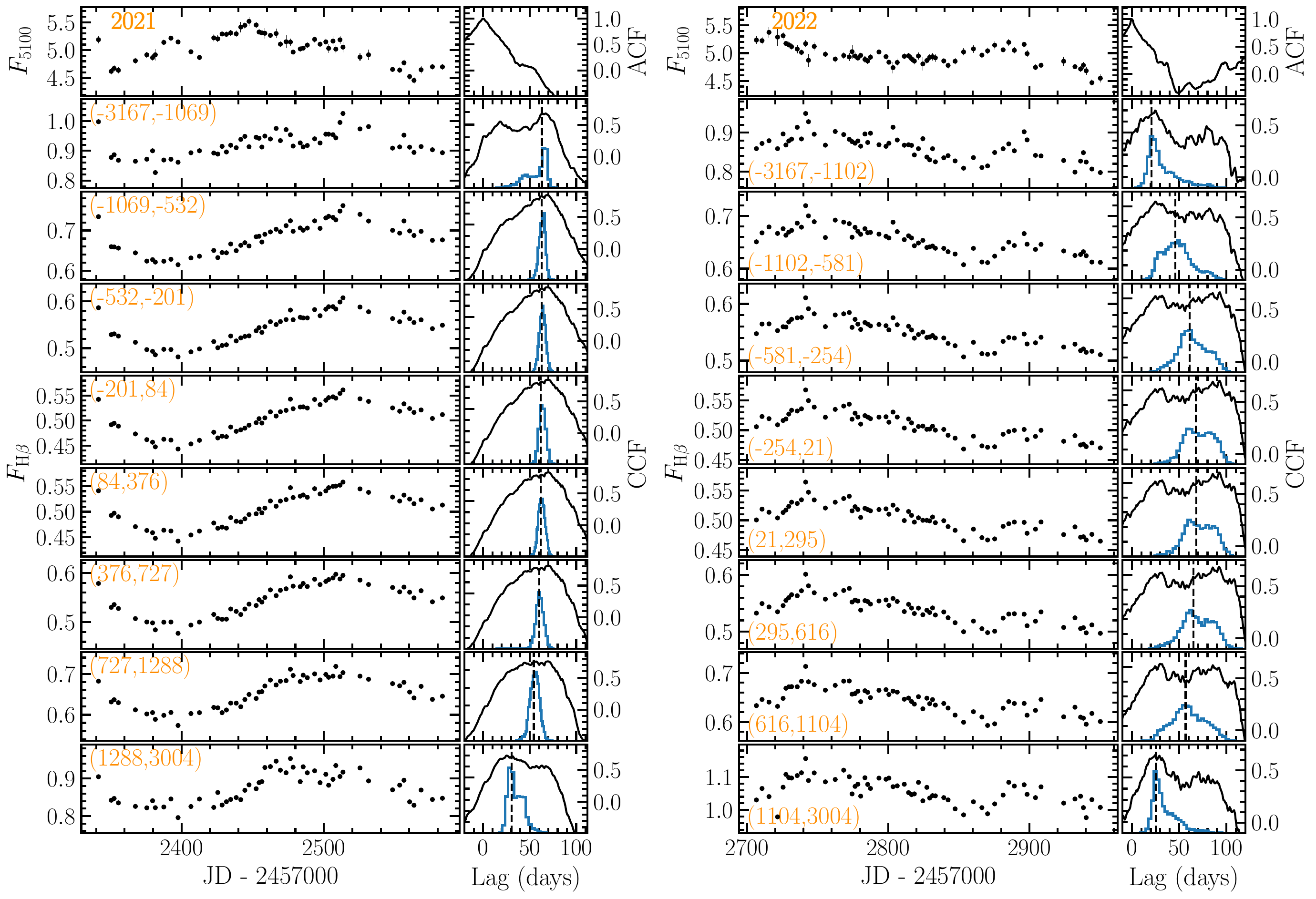}
    \caption{
    Same as Figure~\ref{fig:bin34}, but for 2021 (left) and 2022 (right).
    }
    \label{fig:bin56}
\end{figure*}

Figures~\ref{fig:bin34} and~\ref{fig:bin56} show the {\hb} light curves in each velocity bin with equal flux in the RMS spectrum and corresponding CCF results relative to {\fcs} in all four years, whose time lags are plotted in Figure~\ref{fig:vrd}.

\section{Long-term continuum light curves from several datasets} \label{app:longtermlc}
\begin{figure*}
    \includegraphics[width=2\columnwidth, trim=0 0 40 0]{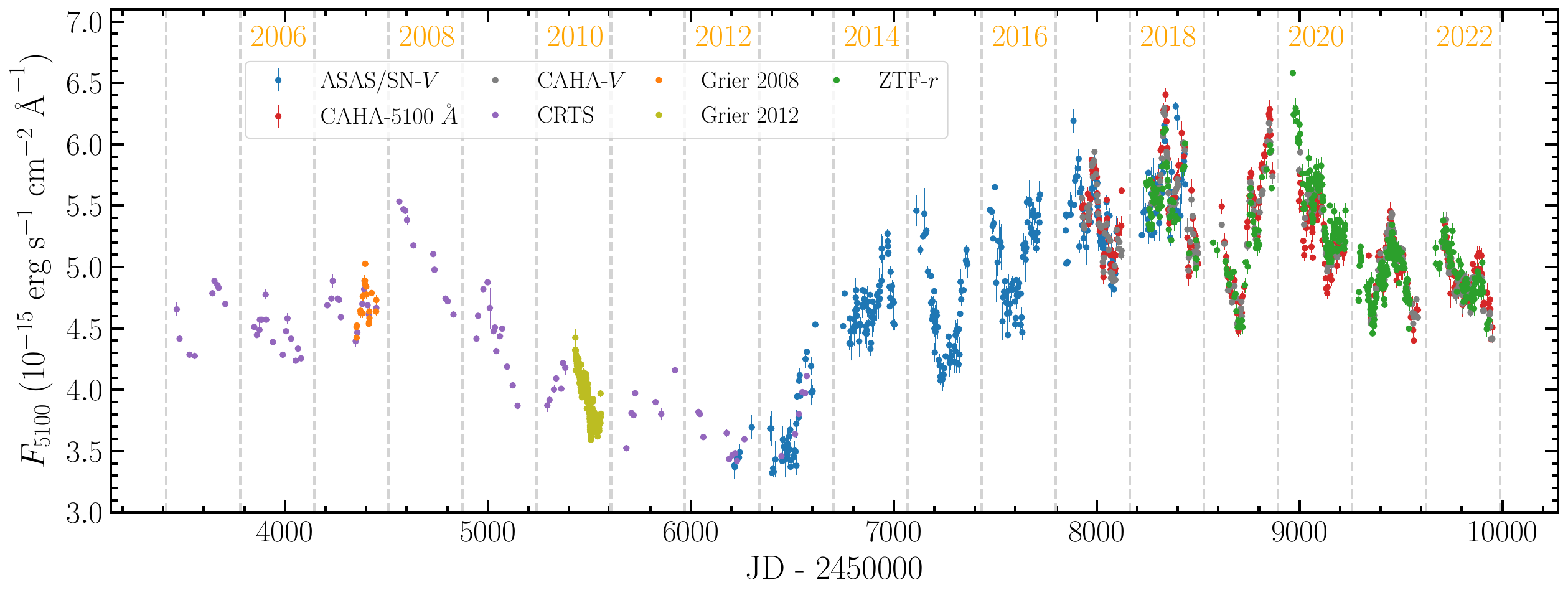}
    \caption{
    Long-term continuum light curves from several datasets adjusted to match the CAHA 5100~{\AA} continuum.
    The gray dashed lines bound the observational seasons (February of the year labeled by the orange text to February of the subsequent year).
    }
    \label{fig:longtermlc}
\end{figure*}

Figure~\ref{fig:longtermlc} shows the long-term continuum light curves from several archival datasets with fluxes intercalibrated to our spectral 5100~{\AA} continuum.

\bibliography{pg2130}{}
\bibliographystyle{aasjournal}

\end{document}